\documentclass[times,twocolumn,preprint]{elsarticle}

\usepackage{medima}
\usepackage{framed,multirow}
\usepackage{rotating}
\usepackage{booktabs}
\usepackage{multirow}
\usepackage{csquotes}
\usepackage{amsmath}
\usepackage{comment}
\usepackage[super]{nth}
\usepackage{diagbox}
\usepackage{textcomp}
\usepackage[dvipsnames]{xcolor}

\usepackage{amssymb}
\usepackage{latexsym}

\usepackage{url}
\usepackage{xcolor}
\usepackage{colortbl}
\usepackage{tabularx}
\usepackage{nicematrix}
\usepackage{subcaption}

\usepackage{hyperref}
\usepackage{cleveref}
\newcolumntype{M}[1]{>{\centering\arraybackslash}m{#1}}
\newcommand*\rot{\rotatebox[x=3.8cm]{90}}
\definecolor{kit}{cmyk}{0.3,0.01,0.18,0}
\definecolor{newcolor}{rgb}{.8,.349,.1}

\journal{Medical Image Analysis}

\begin{document}

\verso{Wachter Andrease \textit{et~al.}}

\begin{frontmatter}

\title{Workflow Augmentation of Video Data for Event Recognition with Time-Sensitive Neural Networks}%

\author[1]{Andreas \snm{Wachter}\corref{cor1}}
\cortext[cor1]{Corresponding author: 
 Tel.: +49-721-608-42650; 
 fax: +49-721-608-42789;
 {E-Mail}: publication@ibt.kit.edu}
\author[1]{Werner \snm{Nahm}}

\address[1]{Karlsruhe Institute of Technology, Kaiserstraße 12, Karlsruhe 76131, Germany}

\received{1 May 2013}
\finalform{10 May 2013}
\accepted{13 May 2013}
\availableonline{15 May 2013}
\communicated{S. Sarkar}


\begin{abstract}
Supervised training of neural networks requires large, diverse and well annotated data sets. In the medical field, this is often difficult to achieve due to constraints in time, expert knowledge and prevalence of an event. Artificial data augmentation can help to prevent overfitting and improve the detection of rare events as well as overall performance. However, most augmentation techniques use purely spatial transformations, which are not sufficient for video data with temporal correlations.\\
In this paper, we present a novel methodology for workflow augmentation and demonstrate its benefit for event recognition in cataract surgery. The proposed approach increases the frequency of event alternation by creating artificial videos. The original video is split into event segments and a workflow graph is extracted from the original annotations. Finally, the segments are assembled into new videos based on the workflow graph.\\
Compared to the original videos, the frequency of event alternation in the augmented cataract surgery videos increased by 26\,\%. Further, a 3\,\% higher classification accuracy and a 7.8\,\% higher precision was achieved compared to a state-of-the-art approach.\\
Our approach is particularly helpful to increase the occurrence of rare but important events and can be applied to a large variety of use cases.
\end{abstract}


\end{frontmatter}



\section{Introduction}
In recent years, deep learning has shown excellent results in medical image analysis and event recognition. Therefore, it is a very promising method to support physicians in their diagnostic and clinical daily routine. Today, deep learning applications are already assisting physicians in diagnosis \citep{Esteva-2017-ID16382}, image registration \citep{Haskins-2020-ID16402}, multi-modal image analysis \citep{Alam-2019-ID16403}, and image segmentation \citep{Hesamian-2019-ID16404}. The most common type of deep learning network is the convolutional neural network (CNN) \citep{Litjens-2017-ID16401}. CNN are engineered to extract information from an image itself using multiple convolutional kernels. However, in the recent years CNNs have also been used for event recognition in surgical workflows from surgical videos \citep{Hajj-2017-ID16202, AlHajj-2019-ID16406, Twinanda-2016-ID16409, Sahu-2016-ID16408, raju2016m2cai}. Event recognition is typically done through the detection of surgical instruments in the images. Nevertheless, the CNNs used still only take into account the image information of individual frames without including the information from the chronological sequence of events. Nevertheless, in most cases a surgical procedure usually follows a predetermined established workflow and therefore, certain instruments appear in the various phases of the workflow in different frequencies or in different sequences. As context recognition has increasingly come into focus in surgery \citep{Pernek-2017-ID16411,Loukas-2018-ID16412}, the recognition of the chronological sequence of events becomes of great importance. Therefore, in the following we answer the question of how neural networks can also make use of such additional temporal information.

\cite{Morita-2019-ID16349} used for surgical phase recognition in cataract surgery the Inception V3 model which is based on a CNN. \cite{Twinanda-2016-ID16410} extended the CNN by a hierarchical hidden Markov model (HHMM) for detecting the individual phase of laparoscopy. Hereby, the CNN extracts the image features and the HHMM considers all temporal information. Twinanda et al. also mention that HHMM is trained separately from the CNN, i.e. EndoNet, due to less training data. \cite{Hajj-2017-ID16202} were able to demonstrate that instead of a combination of CNN and HHMM, the combination of CNN and recurrent neural network (RNN), i.e. long short-term memory (LSTM) network, can considerably increase the recognition performance. 
However, both approaches from \cite{Hajj-2017-ID16202} and \cite{Twinanda-2016-ID16410} still suffer from very small and highly unbalanced training data sets. The authors found out that the precision for the detection of a specific tool is highly correlated with the prevalence of the tool in the training set. This means that in this case we also encounter the common problem of many applications of deep learning in medicine. Frequently, medical data sets are too small and unbalanced for effective training. The consequences are insufficient detection rates and lack of robustness of the neural network, especially for rare events.\\
We therefore conclude that the availability of sufficiently large and adequately balanced training data sets are a necessary prerequisite for the use of deep neural networks for event recognition in the context of surgical workflows. Which leads directly to the scientific question for this paper: How can existing surgical video data sets be retroactively enlarged and balanced, specifically with regard to the frequency and the chronological sequence of events?

\begin{figure}[ht]
 \centering
 \includegraphics[width=\columnwidth] {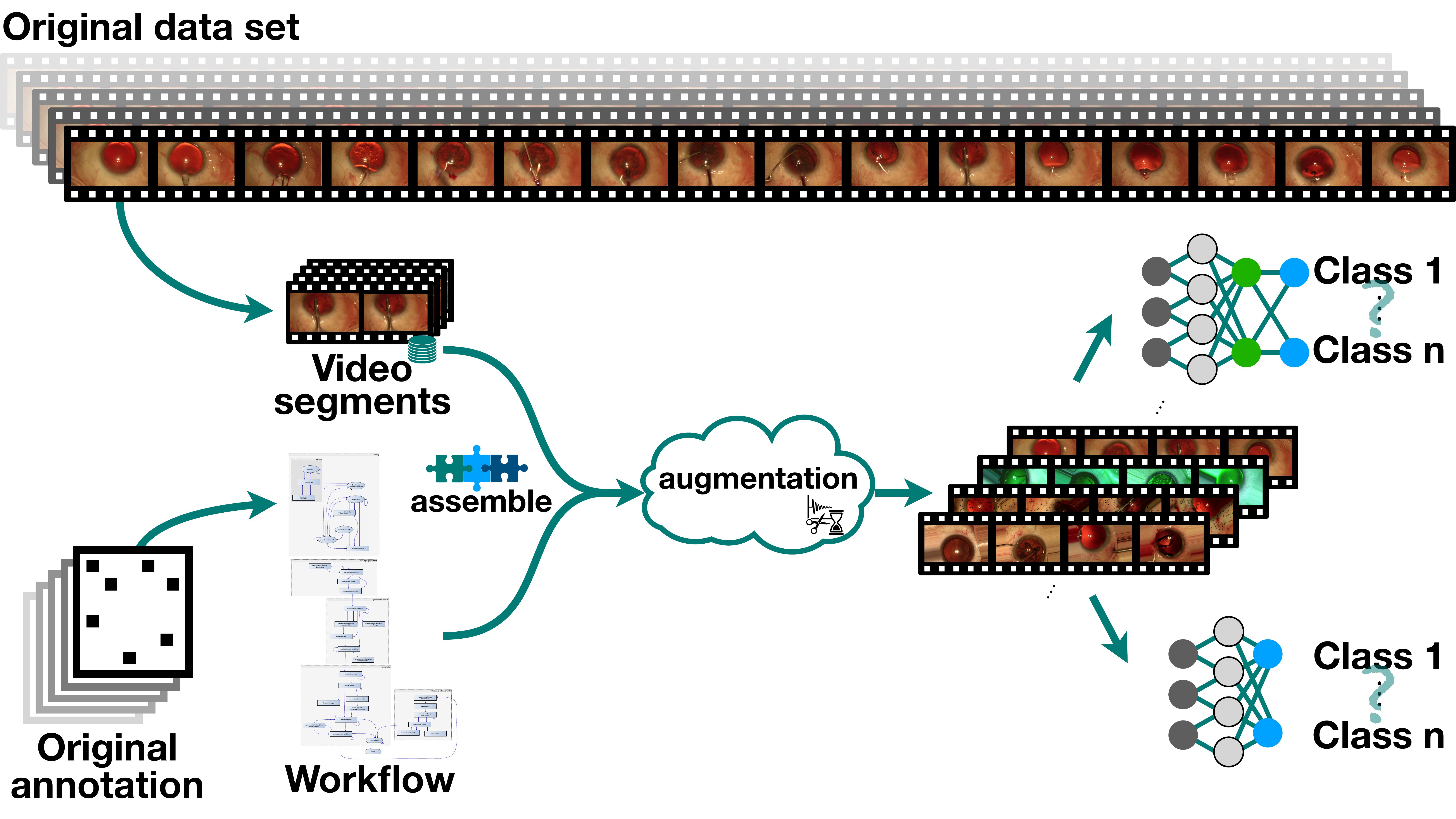}
 \caption{Overview of our approach to augmented time series image data. We extract the workflow from the annotation of original training data set and split the videos along their classifications. Afterwards by using the workflow graph and the sequences we assemble new artificial videos. Additionally they are augmented in spatial and time domain space. In the end we train two different type of neural networks.
 }
 \label{fig:Abstract_image}
\end{figure}

In this study, we present a novel end-to-end workflow-based approach for augmenting and balancing surgical videos. 
In contrast to previous approaches \citep{Hajj-2017-ID16202, AlHajj-2019-ID16406, Lecun-ID16383}, we propose a combination of several methods whose starting point is the extraction of the surgical workflow. Then using this workflow, the artificial videos are reassembled and augmented in both space and time, as shown in~\Cref{fig:Abstract_image}. This methodology allows creating new artificial videos that appear to the neural network as an original recorded video. Furthermore, it is possible to create sequences of the same duration, with variation of speed, and to balance the classifications within the data sets a posteriori.
To the best of our knowledge, there is no approach so far that can enlarge video data sets so that not only the spatial information is augmented, but also the temporal information, and to further balance the data set.

In addition, we will demonstrate that with our method it is possible to augment and balance a cataract video dataset and train a combined CNN and LSTM network whose image classification performance is better than that of the same neural network trained on a dataset augmented using the state-of-the-art method.

This paper is organized as follows: \Cref{sec:related_work} provides an overview of the cataract data set and the current review of the state of the art augmentation methods. We present our methods and the state-of-the-art method which we used for augmenting in \Cref{sec:method}. Additionally, we describe in this section the networks and training strategies which were used. The results are presented in \Cref{sec:results}. The paper ends with a discussion and conclusions in \Cref{sec:discussion} and \Cref{sec:conclusion}, respectively.

\section{Related work}
\label{sec:related_work}
In the following, we describe the current state-of-the-art for augmentations in general, time-series and medical data, as well as the state-of-the-art for the classification of time-series data in medicine. We also highlight the differences in previous work compared to our study.

\cite{Lecun-ID16383} were the first who used data augmentation, i.e. data wrapping, for handwriting recognition using a CNN. Nowadays, data augmentation is already a well-established technique for image recognition. Most of these augmentations are based on random spatial image manipulations, such as geometric, color or noise transformations. Many CNN architectures used different types of these transformations. For example, AlexNet by~\cite{Krizhevsky09} used clipping, mirroring, and color augmentation. AlexNet produced excellent classification results on the ImageNet Large Scale Visual Recognition Challenge (ILSVRC) data set \citep{Russakovsky-ID16389}. Other transformation such as scaling and cropping was used in the Visual Geometry Group (VGG) by~\cite{SimonyanZ14a}, or scaling, cropping and color augmentation in the Residual Networks (ResNet) by~\cite{He-2016-ID16391}, or translation and mirroring in the DenseNet by~\cite{Huang-2017-ID16392}, or cropping and mirroring in the Inception network by~\cite{Szegedy-2015-ID16393}. These transformations are applied separately to the individual images. For those applications the temporal information was not relevant. 

In contrast, there are still only few approaches published for the augmentation of sequential image data such as videos. For video data augmentation \cite{Hajj-2017-ID16202, Parmar-2016-ID16425} suggest a sub-sampling of the original sequence. This increases the size of the data set, but the workflow of the videos always has the exact same time sequence. \cite{Kim-2020-ID16426} suggest for a temporal variation to skip some video frames. This ultimately results in shorter video, which can lead to an unwanted bias of the neural network. To avoid this, \cite{Ji-2019-ID16424} suggests a time warping approach. In this case, the data are scaled in the time domain. Nevertheless, there is a major drawback that Ji et al. applied the method to the extracted feature vector of the input data and not to the original data set. For such an approach, however, the feature data must be unambiguously human-interpretable, which is rarely feasible. Therefore, an end-to-end augmentation approach that augments both the spatial, and temporal information in the video data is desirable.

\section{Methods}
\label{sec:method}
Our approach based on the workflow extracted from meta information in the video. This meta information, i.e. workflow, is present in the data but not limited to them. Using this meta information together with the extracted video segments, we are able to create a balanced data set. This is one of the biggest advantages of our approach. To avoid that the augmentation using workflow is not a duplication of data sequences, we apply spatial data augmentation on each complete video. This is a effective method to enhance the size and quality of the training data, data augmentation is crucial to successful application of deep learning models, when only insufficient data are available. In addition, we apply temporal augmentation by using optical flow not for feature extraction like \cite{Ji-2019-ID16424}, instead to generate sub-frames as \cite{Niklaus_ICCV_2017} suggested. 
The application of optical flow allowed us to also use time warping in the image domain, and not only in the feature domain, as done by \cite{Ji-2019-ID16424}. Such an end-to-end augmentation for time series data, i.e. videos, which is easily human interpretable has never been proposed before.

\begin{figure*}[ht]
	\centering
	\includegraphics[width =\textwidth]{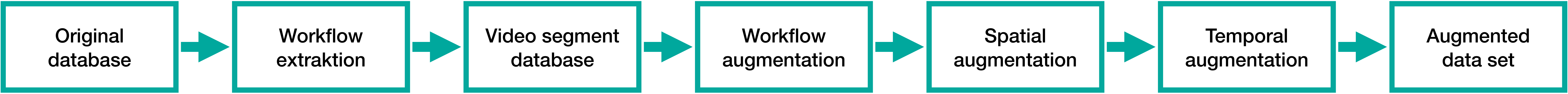}
	\caption{Flow chart of the workflow-based augmentation methodology pipeline.}
	\label{img:blockdiagramm}
\end{figure*}

\subsection{Database}
The goal of this study is to demonstrate the potential of our workflow augmentation method for enlarging and balancing a data set. Therefore, we chose the cataract data set \citep{dataset_cataract}, because it has just a few videos and is highly imbalanced. Furthermore, the possible tool set is limited due to high standardization of the cataract surgery, which reduces the dimensions of the workflow graph. 
We believe that the selection of the cataract data set as an example does not limit the power of our method. A schematic overview pipeline of our approach is shown in \Cref{img:blockdiagramm}. 

 The data set contains 50 cataract surgeries conducted at university hospital Brest. For each surgery, a microscopic video and a surgical tray video was provided for the complete surgical procedure. The surgical video was recorded with an ZEISS OPMI Lumera T microscope (Carl Zeiss Meditec AG, Jena, Germany). The videos are stored as a sequence of images, with an average duration of 10 minutes and 56 seconds. The image format is 1920$\times$\,1080 pixels and the frame rate was approximately 30\,frames per second (fps). The video of the surgical tray, is provided in parallel, as a synchronous side-by-side recording of the surgical tray to the microscope video. Since the tray videos are not used for this investigation, a more detailed description is omitted and can be found in \cite{dataset_cataract}.\\
Additionally, the authors also provide separately for each video the frame-by-frame annotation files for 21 different surgical tools that are in or out of use. The annotation was independently done by two experts with regard to the usage of a tool. Here, usage is defined by whether or not the tool is in contact to the eyeball in the respective frame. The mean of the two experts annotations is used as the final frame annotation. Finally, the annotation of both surgeons was corrected regarding the used tool but not to the exact time of use. Therefore, a certain uncertainty in the annotation of the exact point in time must be taken into account. The full data set was divided into a training set and a test set of 25 videos each by Al Hajj et al.. This training data set is the starting point for all further steps of our approach.

\subsection{Augmentation Method}
In the following, we describe in chronological order our end-to-end workflow augmentation methodology for generating new artificial videos. As a first step we explain how we extract the basic workflow from the existing data set. Next we describe how we create our modular construction kit of video segments and how we generate new videos from this kit using the workflow as a sequence template. Then we describe our exact procedure for varying the generated videos in appearance and temporal dimension Afterwards, we describe the state-of-the-art method, the split augmentation, that we use for comparison.

\subsubsection{Workflow augmentation}
\begin{figure}
 \centering
 \includegraphics[width= 0.9\columnwidth]{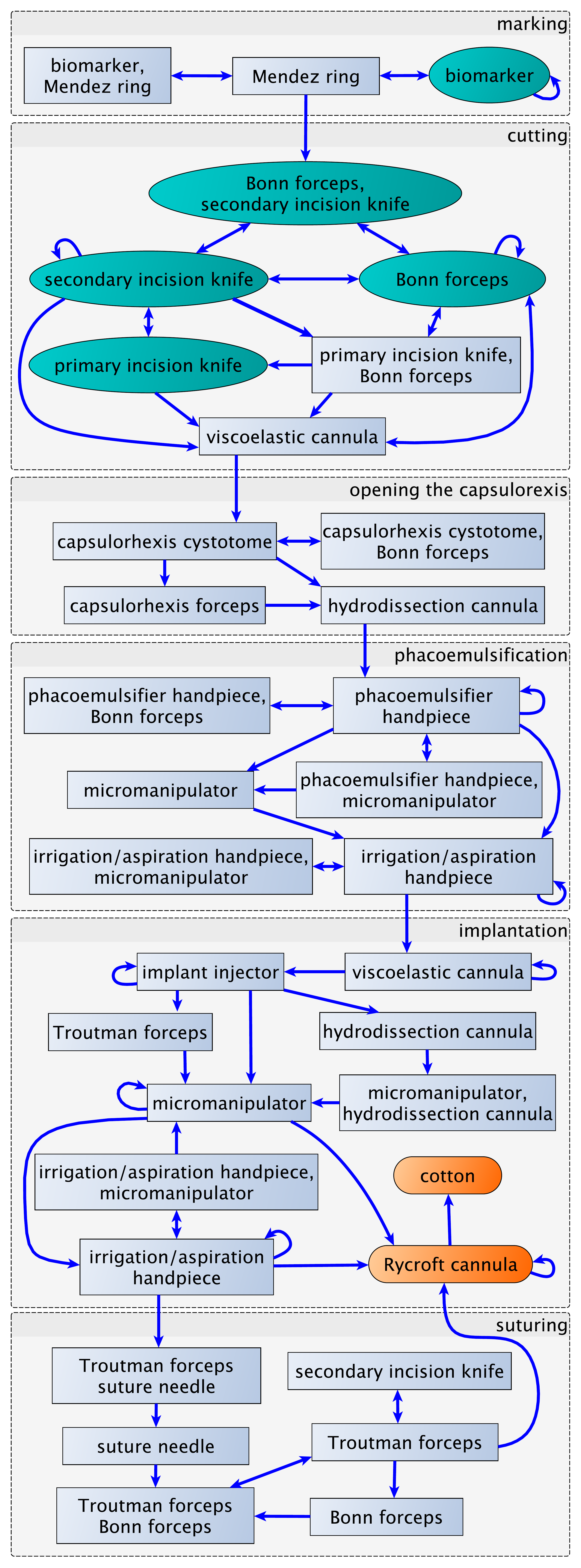}
 \caption{The workflow, which we extracted from the annotation of training data set of the cataract data set. In green marked the starting classes, in orange the final classes and in square boxes the other detected classes. The gray boxes shows the six different phases (marking, cutting, capsulorhexis, phacoemulsification, implantation and suturing) that we identified from the data set. The arrows represent the transitions between the classes. We excluded the following videos: \textit{4}, \textit{8}, \textit{12}, \textit{14}, \textit{19} and \textit{25} due to complexity and testing. For a better overview, the idle intervals are not shown explicitly. }
 \label{fig:finalworkflow}
\end{figure}

\paragraph{\textbf{Workflow extraction}}
\label{para:Workflow_extraction}
In order that the artificial cataract operations videos mimics a real cataract surgery, we require the workflow of a cataract operation as a template. A workflow is an orchestrated and repeatable sequence of work steps. These work steps correspond in our case to typical phases of a surgical procedure. A phase of a surgical procedure always refers to a specific goal or general activity, e.g. opening the surgical cavity. We define a phase as a higher-level action of a surgery. When we look at a surgical phase at the next level of detail, we see a sequence of basic actions. In our example, these actions correspond to the events to be classified, namely the use of surgical tools. The beginning and the end of an event is defined by a change in tool annotation. In the following, the identified events are also named classes. The title of the class indicates the respective tools.\\
In the literature various studies on the procedure of cataract surgery can be found. However, these workflows varied substantially from each other. For example, \cite{Morita-2019-ID16349} propose only a three phase workflow that contains only the essential steps. \cite{Yu-2019-ID16351} propose eight phases, which include two additional case differentiation. \cite{DeepPhase}, their study based on the same data set as ours, propose 14 phases. However, \cite{dataset_cataract} suggested 18 phases for the same data set. But since the phase annotation are not available, we decided to extract the workflow semi-automatically from the annotation files of the training data set, inspired by \cite{Morita-2019-ID16349}, \cite{Yu-2019-ID16351} and \cite{DeepPhase}.\\
Our extracted workflow has 6 unique phases. These phases are: marking, cutting, capsulorhexis, phacoemulsification, implantation and suturing. The phases are linked by specific transitions, i.e. a specific sequence of events and characterized by different anatomical situations and environmental conditions. 
Even if identical tools are used in different phases, this leads to a different context in which the tools are used. Consequently, these tools, even if they are identical but used in a different phase, are later considered as different tools and as different classes, respectively. By this, an ambiguity of the tool classification of the neural network should be prevented.\\
 In consequence, for the extraction of the workflow we first identified all classes that are contained in the full data set. This has been done only based on the annotation files. The sequence of classes varies in the 25 videos. The mean length of a video is 27 classes with a standard variation of $\pm6.5$. The maximum length is 130 (video \textit{19}) and the minimum is 18 (video \textit{23}). The individual sequences of classes were sorted according to length. Afterwards, they were manually merged to an overall workflow. This step could be done automatically as well, but further investigation would be necessary. However, this would go beyond the scope of this paper.\\
To reduce complexity, we decided to exclude very specific surgical phases from the workflow construction, such as IOL removal, which occurs only in the \textit{12} and \textit{14} video. And even if the suture needle appeared to be used in a number of different ways, only one variant was taken into consideration for the workflow. Video \textit{4} was excluded for testing purposes. In the end, we completely excluded the following videos: \textit{4}, \textit{8}, \textit{12}, \textit{14}, \textit{19} and \textit{25}. We are aware that phases, and the tools the excluded files contain, may not be classified correctly during testing. But we assume that this simplification will not limit the methodology. A more complex workflow model would only increase the size of the training data set and thus the training time.\\
Furthermore, the graph of the final workflow is shown in \Cref{fig:finalworkflow}. We identify out of the 25 videos just four starting classes marked in green in round boxes (\textit{biomarker}, \textit{Bonn forceps}, \textit{primary incision knife}, \textit{secondary incision knife}) and two final classes (\textit{cotton}, \textit{Rycorft cannula}) marked in orange round boxes. The squared boxes are intermediate classes. The arrows represent the transitions between the classes.

\begin{figure}[ht]
	\centering
	\includegraphics[width=\columnwidth]{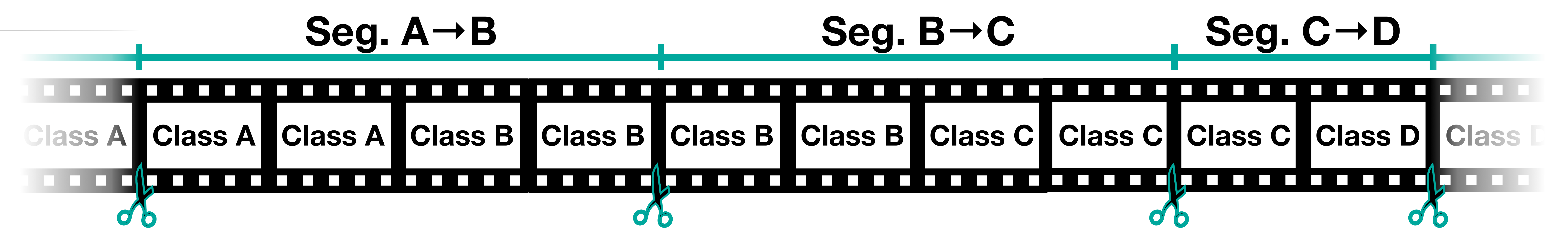}
	\caption{Illustration of splitting the original videos in different segment along the classes.}
	\label{fig:seg_cut}
\end{figure}

\paragraph{\textbf{Video segment database}}
\label{sub:sec:videosegmentDatabase}
For the assembly of the artificial cataract surgery video, the workflow and the video segments are required. Therefore, the videos of the data set have to be split into segments and sorted according to their classification. Before splitting the videos into segments, we had to decide whether the segments should start exactly with the annotation or a few frames before. The best case would be if the segment of a class starts and ends with frames of the class \textit{no tool in contact}. Unfortunately, this is not possible in the cataract data set because, for example, if in the current segment a surgical tool is annotated and another one is added, then the class changes without having the classification \textit{no tool in contact} in between. Therefore, we decided to split the video in the middle of the previous and the current class, as illustrate in \Cref{fig:seg_cut}. This has the advantage that the neural network is not biased by a high entropy due to a cut in the image sequence. More precisely, the neural network should not learn to recognize any cut in the generated video, instead it should interprets the cuts as noise and does not react sensitively to it, because the classification does not change. In addition, it provides more opportunities for a variety of segments combinations in the artificial videos. \\
The segments are stored in a database with their frames-by-frames annotation. The current phase is also saved, allowing instruments that are used multiple times in the workflow can be assembled according to the specific content. In order to have a higher diversity in the database, we included all videos that were excluded during the creation of the workflow, with the only exception of the video \textit{4}. This video was excluded to have an original clinical test video with annotation, because at that time the annotations of the 25 test data from the cataract data set were not public available.\\
As a result for the segment database, we obtain 124 different segments representing an individual type of class transitions. The segments with initial class \textit{no tool in contact} to a class with one or more tools in contact is the largest subset with 45 segments, and vice versa is the second largest subset with 35 segments. 44 segments contain just one individual segment. The segment \textit{Rycroft cannula implantation $ \rightarrow$ no tool in contact} contains the largest subset with 127 different segments. In mean a class contain approx. 8.78 individual segments. The number of frames per segment is in median 184.5\,frames with a the median absolute deviation of 253.2\,frames. The minimal number of frames per segment is 1 and the maximal number are 3733.

\paragraph{\textbf{Workflow augmentation}}
Based on the workflow graph and the video segment database the new artificial cataract surgical videos was created in the next step. The workflow augmentation allows to generated videos have not existed before in this class combination. The initial and final classes are equivalent to one of the original videos. Furthermore, the transition probabilities between the classes has to be determined. They can be chosen arbitrarily, or according to the original data set, which would lead in case of the Cataract data set to an unbalanced enlargement or if they are chosen uniformly distributed to a balanced enlargement. Therefore, we decided to choose the transition probabilities uniformly distributed, because we want to get a balanced data set. We chose the outgoing transition probabilities of each class $w_i$ that they sum up to one:
\begin{equation}
 1 = \sum^E_{i=0} w_i
\end{equation}
In order to prevent \enquote{endless} cycles of a class in the generated artificial videos, the corresponding transition probability was reduced after selection and the other outgoing transition probabilities was increased simultaneously. How strong the decrements and increments should be can be determined by the user. We decided to reduce the probability of the selected transition $w_j$ by $1/2$ and increase probability of the other outgoing transition $w_i$ of the class by difference that they sum up to one in the end:
\begin{equation}
 1 = \frac{w_j}{2}\sum^{E}_{\substack{{i=0},{i \neq j}}} \frac{w_j}{2\cdot(N-1)}+w_i 
\end{equation}

To generate a new video sequence, the transitions of the workflow graph were selected according to their probabilities. In the next step, this class sequence and the segment data base are used to create the corresponding video. If there are more than one segment for one class, the concrete segment in database would be uniformly and randomly selected. Since only tool-\,tool transitions are taken into account in the workflow graph, but the database also contains sequences with the intermediate step \textit{no tool in contact}, we check in advance which variants are available and then choose randomly. As a result, we have now a balanced training data set of artificial videos from an unbalanced original data set, but the segment variation is still small. 

\paragraph{\textbf{Spatial augmentation}}
\label{para:Spatial_augmentation}
Since the produced videos can be very similar, especially if the database contains only a few sequences, we additionally decided to spatial augment the videos. Therefore, we used 15 different augmentation types: center-cropped, padding, rotating by 90$^\circ$, mirroring, zooming, rotating by $\pm90^\circ$, contrast changing, brightness additive or multiplicative changing, gamma-spreading, linearly down sampling, adding Rician noise, adding Gaussian noise, adding Gaussian blur, and adding square noise of the batch generator of \cite{batchgenerator}. In addition, we used two further types of color augmenting: inverting the colors, and randomly shuffling the color channels. We expect that color information in the data, e.g. iris color, will be less important. Only geometric shapes, e.g. of the instruments or the pupil, have a higher influence on the correct classification.\\
Each of the 17 functions is chosen randomly and equally distributed with probability 33\,\% afterwards the execution order randomly shuffled. This results in a larger variation of the augmentation, since the same functions with the exact same parameters but execute in a different order lead to different results. The parameters for the specific augmentation function can be taken from the \Cref{tab:augmentation_parameter}. The values were determined empirically for the data set, allowing that content of classification was still recognizable by the user. 

\begin{table}[ht]
\begin{center}
\caption{Specific augmentation parameter range for the different functions }
\begin{tabular}{l||cc}
\toprule
function & min & max \\ \hline\hline
center cropping & (840,1080) & (1080,1920) \\ 
padding & (1080,2160) & (1920,3840) \\ 
90 degree rotation & 1 & 3 \\ 
mirror axis & \multicolumn{2}{c}{x,y} \\ 
zoom factor & 0.03 & 1 \\ 
random ratation & -90 & 90 \\ 
contrast & 0.2 & 2 \\ 
additive brightness& -64 & 64 \\ 
multiplicative brightness & 0.5 & 1.5 \\ 
gamma spreading & 0.2 & 2 \\ 
linear downsampling& 0.05 & 2 \\ 
Rician noise& 0 & 20 \\ 
Gaussian noise & 0 & 20 \\ 
Gausian blur& 0 & 7 \\ 
square noise& (0,32) & (0,300) \\
\bottomrule
\end{tabular}
\label{tab:augmentation_parameter}
\end{center}
\end{table}

\paragraph{\textbf{Temporal augmentation}}
After augmenting the appearance of the videos using spatial augmentation methods, the videos look different but the segments have exactly the same duration as the original segments. Since we also take into account the temporal information during training, there is still a risk that the neural network can memorize the few original segments, which leads to over fitting. For this reason we decided to augment the videos also in the temporal domain.\\
A possible augmentation method would be to increase or decrease the duration of the segments by change of the frame rate. This can be realized by dropping or duplicating individual frames. However, these techniques have a significant disadvantage. Dropping or duplicating frames leads to discontinuity of the optical flow in the video. This can be recognized by non-natural jumps, non-smooth movements of the surgical tools. As a result, this could negatively affect the performance of the network, because the videos would have an unnatural character and the inter frame differences are zero in case of duplicating and further to no new or various information.\\
For this reason, we decided to use the optical flow to generate new sub-frames. For generating a sub-frame from two input frames, we used the implementation of \cite{Niklaus_ICCV_2017}. Niklaus et al. proposed an approach using an encoder-decoder network that extracts features out of two given input frames. These features are the inputs of four sub-neural networks. Each of these sub-networks estimate an one dimensional kernel for each output pixel in a pixel-wise manner. Afterwards the estimated kernels are convoluted with the two input frames to obtain the interpolated.\\
For the temporal augmentation range, we chose 20 different factors for speed variation. These are divided equally between $0.5{\times} - 1{\times} - 2{\times}$ and have the following values: 2, 1.8824, 1.778, 1.6842, 1.6, 1.4884, 1.3913, 1.3061, 1.2075, 1.1637, 1, 0.9552, 0.9014, 0.8533, 0.8, 0.7529, 0.7033, 0.6534, 0.5981, 0.5517 and 0.5. These speed factors correspond to the following sub-frames: 128, 116, 107, 98, 91, 85, 80, 75, 71, 67, 64, 58, 53, 49, 46, 43, 40, 38, 36, 34 and 32. Therefore, the sampling rate was increased between two frames by adding 64 interpolating frames. Retrospectively, we have to recognize that the full up-sampling can also be applied directly on the segment data set. This would result in a significant speed up, especially in the case of a large number of videos that have to be augmented.\\
For the annotation of the sub-frames $[1-32]$ the annotation of the original frame $n$ was taken and for the sub-frames $[33-64]$ the annotation of the frame $n+1$. In contrast to the spatial augmentation, we did not augment the entire generated video with one parameter set, instead we divided the video into randomly long parts according to the mean plus/minus the mean absolute deviation of the original class sequences. Then we augmented the individual parts of the video with a randomly selected discrete speed factor. We expect this will lead to greater variability in duration of individual surgical processes and further to more classification robustness of the neural network. To ensure that the complete range for the augmentation length for a video and speed variations is covered as uniformly as possible, we created a Halton sequence. We chosen the Halton sequence, more precisely a two-dimensional Halton sequence from $[0-1]$, because it has a small discrepancy, i.e. the sequence is randomly, and covers the complete range uniformly. Afterwards we linear interpolated the samples to that the first dimension represents the sequence length and the second dimension the speed factor. The concrete values for are randomly selected from the Halton sequence.\\
This was the last step of our workflow augmentation approach and as a result we got new artificial videos from the data set. These videos are correlated the original videos, but they are more balanced.

\subsubsection{Comparison approach: split augmentation}
\label{subsubsec:split_augmentation}
\cite{Shen-2017-ID16525} and \cite{Hajj-2017-ID16202} introduced an approach to artificially increase the samples of sufficient data set. Thereby, Al~Hajj~et~al. split the long videos of the cataract surgery into smaller subset. They split a original video into ten sub-videos by taking just every \nth{10} frames of the video.\\
We decided to use this approach as a comparison method. This approach allows to artificially increase the very small amount of data without including other videos or manipulate the content of the videos. In addition, since the individual sub-videos do not differ much from each other, we decided to spatial augment each sub-videos individually, that they appears different. For the spatial augmentation we chose the same procedure and parameters as described in \Cref{para:Spatial_augmentation},~Spatial augmentation. Furthermore for better comparability, we chose only the videos that were used to create the workflow graph (see \Cref{para:Workflow_extraction},~Workflow extraction), these were in total 19 original cataract videos, which results into 190 sub videos. 

\subsection{Training strategy}
\label{subsec:training_strategy}
For the training of the neural networks on the both data sets, the workflow augmented and the split augmented, we converted the annotation for each image to a binary annotation. Therefore, labels that are greater than 0.5 were set to 1, others to 0. Additionally, we converted the multiple-class annotation into a multi-class annotation. This resulted in 27 different classes instead of 21 classes. Furthermore, we normalized the images over the complete data set. Then, we divided the 5000 videos of the workflow augmented data set into 3000 videos (60\,\%) for training, and 2 $\times$ 1000 videos (20\,\%) for validation and testing, respectively. The exact same sets of data from one data set were used for training and validation of both classifiers. For the split data set, we decided to take all 190 videos for the training and use the 1000 validation videos from the workflow augmented data set for validation due to the small number of independent videos in the split data set. We used the standard stochastic gradient descent (SGD) for optimization with a learning rate of $lr=0.01$ and a momentum of $momentum=0.9$. The learning rate was linearly decreased by $0.1$ every \nth{10} iteration. We trained our models in parallel, on 8 NVIDIA Tesla V100s with 32 GB using module-level data parallelism. Due to memory constraints, only a minimum of every \nth{15} frame, which corresponds to a frame rate of 2\,fps, and a maximum of 2400\,frames per video were selected for the workflow augmented data set for training the LSTM classifier.\\ 
For the CNN classifier, the maximum of 2400\,frames were used whenever possible, randomly selected over the total length of each video and finally randomly shuffled. For the training of CNN and LSTM classifier on the split data set, all frame were used. For the validation, as before for the workflow augmented data set, every \nth{15} frame and maximal 2400\,frames were used. Additionally, the frames of the training set for the CNN classifier were randomly shuffled as before. The order of the training set were randomly selected for both data sets. For the evaluation, we used the cross entropy loss and a one-hot classifier. We trained the models until they converged. 

\subsection{Implementation details of the neural networks}
\label{sec:details}

\paragraph{\textbf{CNN network}}
We selected the convolutional neural network (CNN) ResNet50 by~\cite{resnet50} as a comparison network that takes into account the temporal domain. Because the ResNet50 in combination with other neural networks showed at the cataract challenge~\cite{AlHajj-2019-ID16406} that it was one of the best performing networks. Additionally, we decided to use transfer learning due to the small amount of training data available. We replaced the last fully connected layer with a type-identical layer of 27 nodes representing the 26 binary multi-classes and the class \textit{no tool in contact}. The weight of this layer was randomly initialized. Since the complete data set was normalized, we decided to disable the estimation of batch normalization for each batch normalization layer. The variable track\_running\_stats was set to \textit{False} and the running\_mean and running\_var to \textit{None}. As a result, the scaling factor $\gamma$ is 1 and the shift factor $\beta$ is 0 for each batch normalization layer. The batch normalization was based only on the current batch, which always represents a complete cataract surgery proceeding. The normalization $y$ of image $x$ are calculated with the following equation:

\begin{equation}
 y(x) = \frac{x- E[x]}{\sqrt{Var[x]+ \epsilon}} \text{ ,}
\end{equation}
where $E$ is the expected value, $Var$ the variance and $\epsilon$ are by default 1e-5.
All described parameters with the chosen value are also listed in \Cref{Tab:Resnet}.

\begin{table}[ht]
\centering
\caption{Selected and changed parameters and their values for the training of the network RestNet50}
\begin{tabular}{l||c}
\toprule
\textbf{Parameter} & \textbf{Value} \\ \hline\hline
\begin{tabular}[l]{@{}l@{}}Data set split in \,\% for \\ workflow augmentation\end{tabular} & 
\begin{tabular}[c]{@{}c@{}}60/20/20\\ (train/ valid/ test)\end{tabular} \\ 
Optimizer & SGD \\
Classifier & one-hot classifier \\
Learning rate & $0.01 - 0.1 \cdot \lfloor\frac{epoch}{10}\rfloor$ \\
Momentum & 0.9 \\
Max batch size & 2400 frames \\ 
Step width for LSTM & 15 frames\\
Batch normalization layer: & \\
\hspace{1em} track\_running\_stats & \textit{False} \\
\hspace{1em} running\_mean & \textit{None}\\ 
\hspace{1em} running\_var & \textit{None}\\
\bottomrule
\end{tabular}%
\label{Tab:Resnet}
\end{table}

\paragraph{\textbf{LSTM network}}
To keep the influence of the network selection as low as possible, we extended the previously explained ResNet50 by a layer of 21 LSTM nodes~\citep{Hochreiter-1997-ID15662} and a fully connected linear output layer with 27 nodes. 
To avoid learning the network from scratch and to make the training of the LSTM layer more stable, we used the previously described ResNet50 pretrained on the images over 8 epochs. 

\section{Results}
\label{sec:results}
We conducted three experiments to evaluate the performance of our workflow augmentation approach. The first experiment is the classification performance of the CNN and LSTM network on the workflow augmented test data set. For the second, we evaluate both network on a real test data set. For the third, we evaluate the split trained networks on the same real test data set. Before we evaluated the experiments, we further investigated the generated data set that were produced by our presented approach.

\subsection{Generated videos of workflow augmented data set}
\label{subsec:res:Workflow_augmentation_data set}
The generated workflow data set contains 5000 artificial videos. They are generated with the workflow graph by using 19 original videos and the corresponding segments from 24 videos.
\begin{figure}[ht]
 \begin{minipage}{0.25\columnwidth}
 \subcaptionbox{}
 {\includegraphics[width=\linewidth]{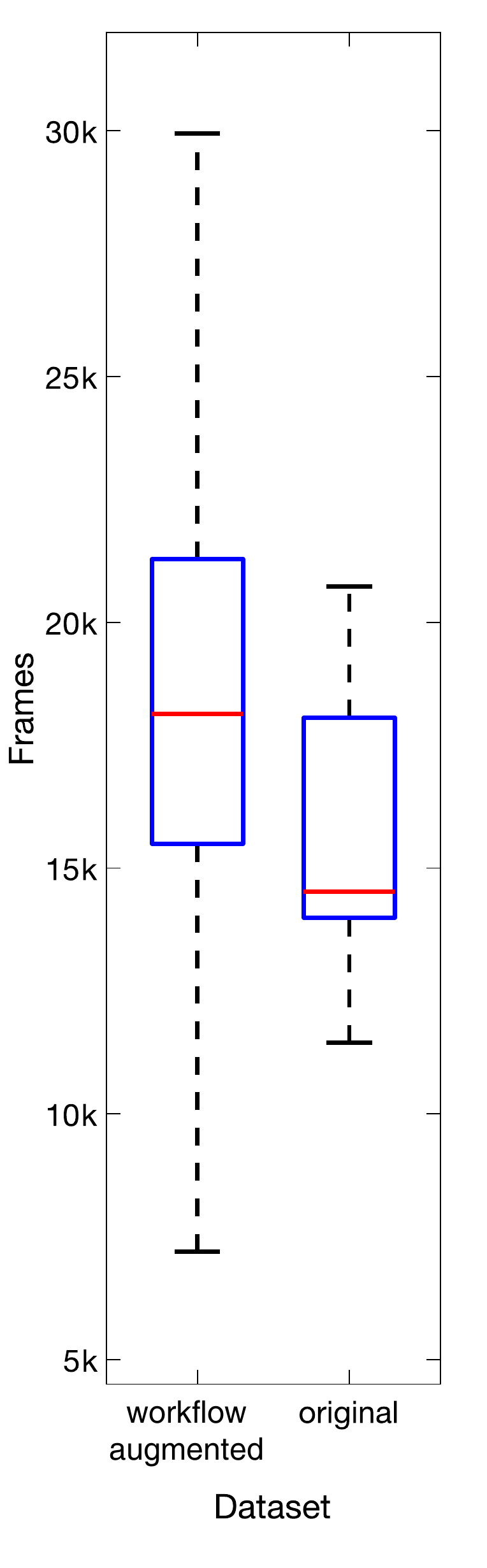}} %
 \end{minipage}%
 \hfill
 \begin{minipage}{0.25\columnwidth}
 \subcaptionbox{}
 {\includegraphics[width=\linewidth]{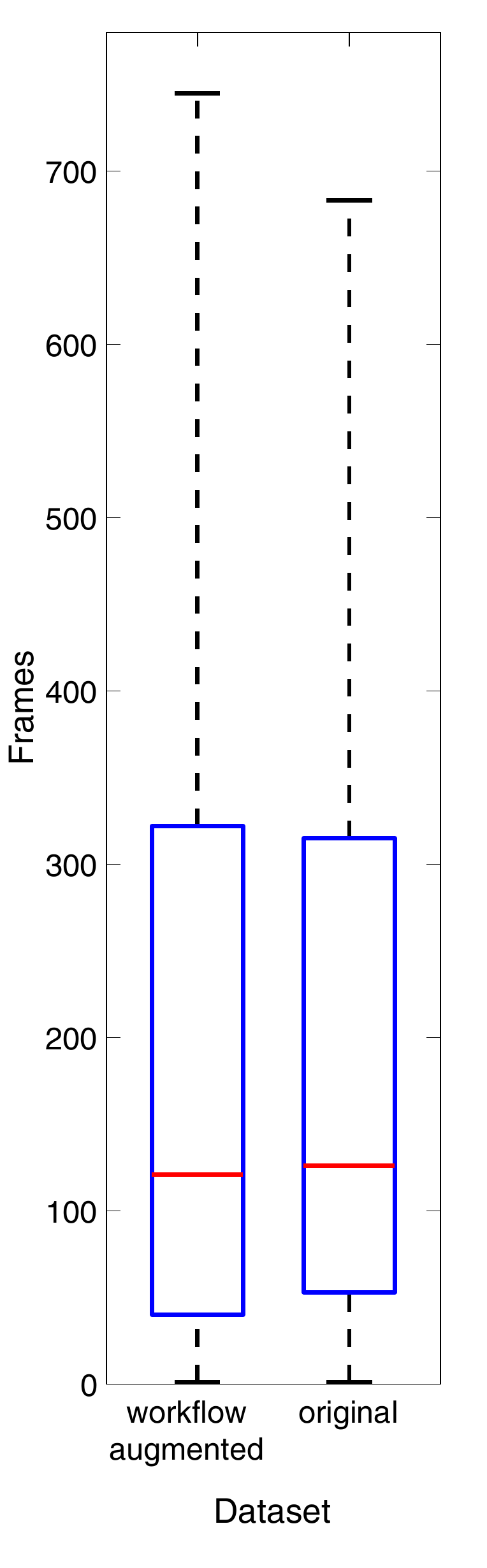}} %
 \end{minipage}%
 \hfill
 \begin{minipage}{0.25\columnwidth}
 \subcaptionbox{}
 {\includegraphics[width=\linewidth]{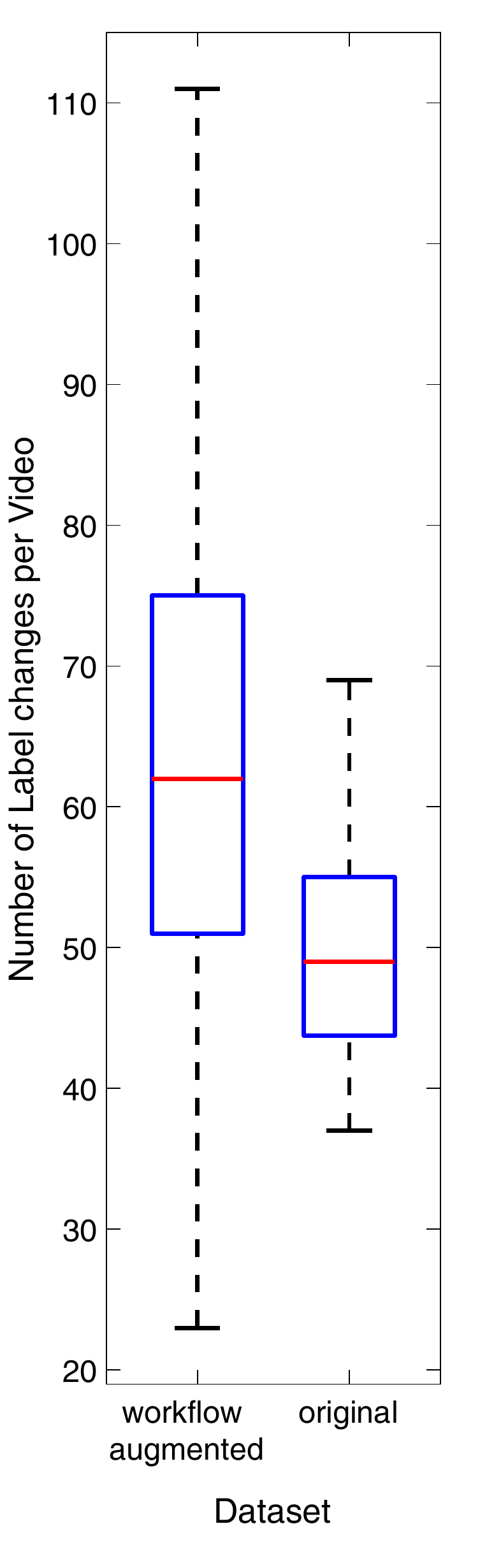}} %
 \end{minipage}%
 \hfill
 \begin{minipage}{0.25\columnwidth}
 \subcaptionbox{}
 {\includegraphics[width=\linewidth]{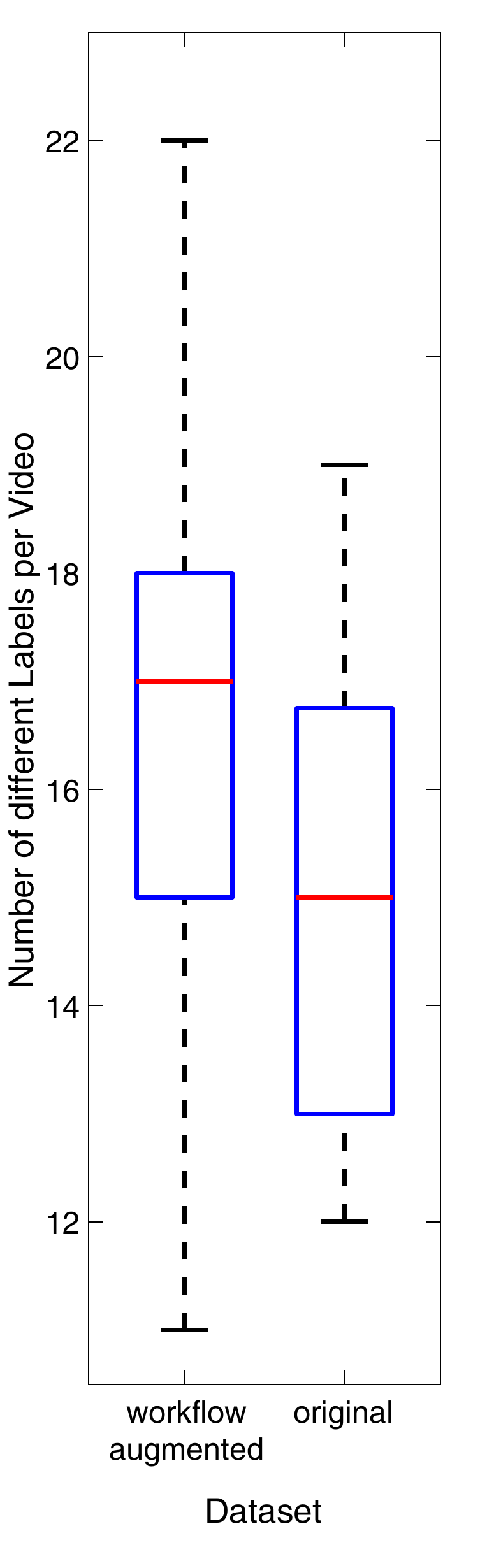}} %
 \end{minipage}%

\caption{Boxplots of indicators of the data set augmented by the workflow (on the left in each subfigure) and the original data set. Each boxplot shows in red the median and in blue the inter quantile distance \nth{25} percentile and \nth{75} percentile. In black dashed line are the whiskers with the \nth{25} percentile minus, resp. \nth{75} percentile plus 1.5 times the inter quantile distance. (a) Shows the plots for the total video length. (b) Shows the plots for the individual sequence length of the videos. (c) Shows the plots for the number of label changes of each video. (d) Shows the plots of the number of different labels for the each video.}
\label{fig:boxplot}
\end{figure}

\begin{table}[h!]
\centering
\caption{Comparison of the data set augmented by the workflow and original data set}
\label{tab:Comparison_workflow_org}
\renewcommand{\arraystretch}{1.2}
\resizebox{\columnwidth}{!}{%
\begin{tabular}{lc||c|c|c}
\toprule
 & & \nth{25} perc. & median & \nth{75} perc.\\ \hline\hline
\multirow{2}{*}{\begin{tabular}[c]{@{}l@{}}Total video \\ length (frames)\end{tabular}} & original & 13993 & 14525 & 18061 \\\cline{2-5}
 & workflow aug. & 15496.5 & 18134 & 21282.5 \\ \hline\hline
\multirow{2}{*}{\begin{tabular}[c]{@{}l@{}}Sequence \\ length (frames)\end{tabular}} & original & 53 & 126 & 315 \\\cline{2-5}
 & workflow aug. & 40 & 121 & 322 \\ \hline\hline
\multirow{2}{*}{\begin{tabular}[c]{@{}l@{}}\# label changes\\ per video\end{tabular}} & original & 43.75 & 49 & 55 \\ \cline{2-5}
 & workflow aug. & 51 & 62 & 75 \\ \hline\hline
\multirow{2}{*}{\begin{tabular}[c]{@{}l@{}}\# diff. labels \\ per video\end{tabular}} & original & 13 & 15 & 16.75 \\\cline{2-5}
 & workflow aug. & 15 & 17 & 18 \\ 
 \bottomrule
\multicolumn{5}{l}{\footnotesize (\#): number of}
\end{tabular}%
}
\end{table}

\begin{table}[ht]
\centering
\caption{Class distribution of the different data sets in percent}
\resizebox{\columnwidth}{!}{%
\begin{tabular}{l|c|c|c}
\toprule
\multirow{2}*{Classes} & workf. & org. & test \\ 
 & augment. & train & set\\ \hline \hline
no tool in contact & 50.069 & 46.304 & 42.687 \\
biomarker & 0.084 & 0.026 & 0.045 \\
hydrodissection cannula & 2.057 & 2.021 & 1.843 \\
Rycroft cannula & 0.734 & 3.201 & 3.382 \\
viscoelastic cannula & 2.766 & 1.614 & 1.549 \\
cotton & 1.417 & 0.147 & 0.010 \\
capsulorhexis cystotome & 4.930 & 4.687 & 5.877 \\
Bonn forceps & 0.209 & 0.184 & 0.028 \\
capsulorhexis forceps & 1.255 & 0.970 & 0.537 \\
Troutman forceps & 0.758 & 0.086 & 0.007 \\
irrigation/aspiration handpiece & 17.197 & 16.769 & 16.479 \\
phacoemulsifier handpiece & 2.410 & 3.274 & 3.999 \\
implant injector & 1.705 & 1.464 & 1.522 \\
primary incision knife & 0.429 & 0.448 & 0.576 \\
secondary incision knife & 0.464 & 0.296 & 0.373 \\
micromanipulator & 2.176 & 1.596 & 1.483 \\
suture needle & 0.110 & 0.027 & 0.000 \\
Mendez ring & 0.448 & 0.162 & 0.133 \\
Mendez ring \& biomaker & 0.003 & 0.001 & 0.014 \\
\begin{tabular}[c]{@{}l@{}}Bonn forceps \& \\secondary incision knife\end{tabular} & 0.320 & 0.292 & 0.255 \\
\begin{tabular}[c]{@{}l@{}}primary incision knife \& \\Bonn forceps\end{tabular} & 0.567 & 0.324 & 0.174 \\
\begin{tabular}[c]{@{}l@{}}capsulorhexis cystotome \& \\Bonn forceps\end{tabular} & 2.781 & 0.361 & 0.028 \\
\begin{tabular}[c]{@{}l@{}}phacoemulsifier handpiece \& \\Bonn forceps\end{tabular} & 0.083 & 0.052 & 0.129 \\
\begin{tabular}[c]{@{}l@{}}phacoemulsifier handpiece \& \\micromanipulator\end{tabular} & 5.426 & 13.336 & 14.441 \\
\begin{tabular}[c]{@{}l@{}}irrigation/aspiration handpiece \& \\micromanipulator\end{tabular} & 1.229 & 2.280 & 4.429 \\
\begin{tabular}[c]{@{}l@{}}hydrodissection cannula \& \\micromanipulator\end{tabular} & 0.243 & 0.039 & 0.000 \\
\begin{tabular}[c]{@{}l@{}}Troutman forceps \& \\suture needle\end{tabular}
 & 0.127 & 0.040 & 0.000 \\ 
\bottomrule
\end{tabular}
\label{tab:tooldistribution}
}
\end{table}

\begin{table*}[ht]
\centering
\caption{Binary analysis of the classification for the CNN and LSTM classifier on the test data of the workflow augmented data set with the scores for accuracy (ACC), precision (Prec), recall (Rec), specificity (Spec), and F1-score, rounded on 4 digits.}
\resizebox{0.95\textwidth}{!}{%

\begin{tabular}{l||M{6ex}M{6ex}M{6ex}M{6ex}M{6ex}||M{6ex}M{6ex}M{6ex}M{6ex}M{6ex}} 
\toprule
& \multicolumn{5}{c||}{CNN} & \multicolumn{5}{c}{LSTM} \\ \cline{2-11} 
 & ACC & Prec & Rec & Spec & F1-score & ACC & Prec & Rec & Spec & F1-score \\ \hline
no tool in contact & 0.9785 & 0.9668 & 0.9907 & 0.9666 & 0.9786 & 0.9750 & 0.9601 & 0.9908 & 0.9595 & 0.9752 \\
biomarker & 0.9997 & 0.9080 & 0.6799 & 0.9999 & 0.7776 & 0.9995 & 0.9385 & 0.4988 & 1 & 0.6514 \\
hydrodissection cannula & 0.9976 & 0.9487 & 0.9307 & 0.9990 & 0.9396 & 0.9974 & 0.9444 & 0.9239 & 0.9989 & 0.9340 \\
Rycroft cannula & 0.9980 & 0.9417 & 0.7879 & 0.9996 & 0.8579 & 0.9977 & 0.9341 & 0.7488 & 0.9996 & 0.8313 \\
viscoelastic cannula & 0.9966 & 0.9461 & 0.9307 & 0.9985 & 0.9383 & 0.9961 & 0.9385 & 0.9200 & 0.9983 & 0.9291 \\
cotton & 0.9981 & 0.9685 & 0.9001 & 0.9996 & 0.9331 & 0.9963 & 0.9820 & 0.7581 & 0.9998 & 0.8556 \\
capsulorhexis cystotome & 0.9963 & 0.9785 & 0.9463 & 0.9989 & 0.9621 & 0.9967 & 0.9682 & 0.9637 & 0.9984 & 0.9659 \\
Bonn forceps & 0.9993 & 0.9143 & 0.7487 & 0.9999 & 0.8232 & 0.9993 & 0.8692 & 0.7665 & 0.9998 & 0.8146 \\
capsulorhexis forceps & 0.9989 & 0.9770 & 0.9408 & 0.9997 & 0.9585 & 0.9993 & 0.9836 & 0.9659 & 0.9998 & 0.9747 \\
Troutman forceps & 0.9990 & 0.9657 & 0.9039 & 0.9997 & 0.9337 & 0.9989 & 0.9671 & 0.8912 & 0.9998 & 0.9276 \\
irrigation/aspiration handpiece & 0.9920 & 0.9776 & 0.9762 & 0.9953 & 0.9769 & 0.9921 & 0.9797 & 0.9745 & 0.9958 & 0.9771 \\
phacoemulsifier handpiece & 0.9975 & 0.9488 & 0.9458 & 0.9988 & 0.9473 & 0.9972 & 0.9652 & 0.9174 & 0.9992 & 0.9407 \\
implant injector & 0.9985 & 0.9707 & 0.9400 & 0.9995 & 0.9551 & 0.9986 & 0.9736 & 0.9435 & 0.9996 & 0.9583 \\
primary incision knife & 0.9991 & 0.9580 & 0.8494 & 0.9998 & 0.9004 & 0.9990 & 0.9317 & 0.8342 & 0.9997 & 0.8803 \\
secondary incision knife & 0.9989 & 0.9221 & 0.8184 & 0.9997 & 0.8671 & 0.9988 & 0.9078 & 0.8116 & 0.9996 & 0.8570 \\
micromanipulator & 0.9971 & 0.9506 & 0.9174 & 0.9989 & 0.9337 & 0.9973 & 0.9439 & 0.9304 & 0.9988 & 0.9371 \\
suture needle & 0.9997 & 0.9479 & 0.7575 & 1 & 0.8421 & 0.9997 & 0.9536 & 0.7258 & 1 & 0.8243 \\
Mendez ring & 0.9996 & 0.9810 & 0.9293 & 0.9999 & 0.9545 & 0.9996 & 0.9823 & 0.9375 & 0.9999 & 0.9594 \\
\begin{tabular}[c]{@{}l@{}}Mendez ring \&\\ biomarker\end{tabular} & 1 & 1 & 0.1951 & 1 & 0.3265 & 1 & 1 & 0.9756 & 1 & 0.9877 \\
\begin{tabular}[c]{@{}l@{}}Bonn forceps \& \\ secondary incision knife\end{tabular} & 0.9994 & 0.9177 & 0.9020 & 0.9997 & 0.9098 & 0.9994 & 0.9395 & 0.8707 & 0.9998 & 0.9038 \\
\begin{tabular}[c]{@{}l@{}}primary incision knife \& \\ Bonn forceps\end{tabular} & 0.9994 & 0.9629 & 0.9356 & 0.9998 & 0.9491 & 0.9993 & 0.9675 & 0.9121 & 0.9998 & 0.9390 \\
\begin{tabular}[c]{@{}l@{}}capsulorhexis cystotome \& \\ Bonn forceps\end{tabular} & 0.9994 & 0.9966 & 0.9818 & 0.9999 & 0.9892 & 0.9996 & 0.9984 & 0.9864 & 1 & 0.9924 \\
\begin{tabular}[c]{@{}l@{}}phacoemulsifier handpiece \& \\ Bonn forceps\end{tabular} & 0.9999 & 0.9596 & 0.9344 & 1 & 0.9468 & 0.9999 & 0.9393 & 0.9134 & 1 & 0.9261 \\
\begin{tabular}[c]{@{}l@{}}phacoemulsifier handpiece \& \\ micromanipulator\end{tabular} & 0.9978 & 0.9881 & 0.9717 & 0.9993 & 0.9798 & 0.9974 & 0.9944 & 0.9592 & 0.9997 & 0.9765 \\
\begin{tabular}[c]{@{}l@{}}irrigation/aspiration handpiece \& \\ micromanipulator\end{tabular} & 0.9984 & 0.9798 & 0.8976 & 0.9997 & 0.9369 & 0.9984 & 0.9739 & 0.9070 & 0.9997 & 0.9393 \\
\begin{tabular}[c]{@{}l@{}}hydrodissection cannula \& \\ micromanipulator\end{tabular} & 0.9999 & 0.9887 & 0.9836 & 1 & 0.9861 & 0.9999 & 0.9882 & 0.9845 & 1 & 0.9864 \\
\begin{tabular}[c]{@{}l@{}}Troutman forceps \& \\ suture needle\end{tabular} & 0.9998 & 0.9549 & 0.9287 & 0.9999 & 0.9416 & 0.9998 & 0.9586 & 0.9068 & 0.9999 & 0.9320 \\
\bottomrule
\end{tabular}
}
\label{tab:bin_eval_aug_test}
\end{table*}

The original videos have a frame count of 14525 in the median with the \nth{25}\,percentile of 13993 and \nth{75}\,percentile of 18061\,frames. The videos of the workflow augmented data set had a length of 18134\,frames in median, with a \nth{25}\,percentile of 15496.5 and \nth{75}\,percentile of 21282.5\,frames (shown in \Cref{fig:boxplot}\,(a) and \Cref{tab:Comparison_workflow_org}). \Cref{fig:boxplot}\,(b) and \Cref{tab:Comparison_workflow_org} shows a similar distribution for the segment lengths for the workflow augmented and for the original data set. The segments in the original videos had a median frame count of 126 with \nth{25}\,percentiles of 53 and \nth{75}\,percentiles of 315\,frames. The segments of the videos of the workflow augmented data set had a median length of only 121\,frames with a \nth{25}\,percentile of 40 and \nth{75}\,percentile of 322\,frames. As a result, the original video are shorter and the length variability are not so uniformly distributed as for the workflow augmented data set. Furthermore, the segment length variability of the original data set and the workflow augmented data set are also very similar.\\ 
The workflow augmented data set had a higher alternation of classifications, with a median of 62 changes and a \nth{25}\,percentile of 51 and \nth{75}\,percentile 75 changes, within a video. The videos in the original data set had in median of 49 class changes per video and a \nth{25}\,percentile of 43.75 changes and \nth{75}\,percentile of 55 changes (shown in \Cref{fig:boxplot}\,(c) and \Cref{tab:Comparison_workflow_org}). Also, the number of different tools and tool combinations used were greater than in original data set with a median of 17 classes and a \nth{25}\,percentile of 15 and \nth{75}\,percentile of 18. In the original data set, the median of the classes is 15 and the \nth{25}\,percentile 13 and the \nth{75}\,percentile 16.75 classes (shown in \Cref{fig:boxplot}\,(d) and \Cref{tab:Comparison_workflow_org}). \Cref{tab:tooldistribution} shows the class distribution of both data set. The table also indicates that the classes of the workflow augmented data set are more balanced than those of the original data set. Most of the classes appeared more frequently in the workflow augmented data set except the few exceptions mentioned below: \textit{Rycroft cannula}, \textit{phacoemulsifier handpiece}, \textit{primary incision knife}, \textit{phacoemulsifier handpiece \& micromanipulator} and \textit{irrigation/aspiration handpiece \& micromanipulator}. Further, the workflow augmentation approach increased the amount of rare class, e.g, \textit{hydrodissection cannula \& micromanipulator} approximately by a factor of 6 from 0.039\,\% of all frames in the original data set to 0.243\,\% in the augmented data set.

\subsection{Classification of the networks on the workflow augmented test data set}
\label{subsec:classification_workflow_test}
The first experiment is designed to test the performance of the classifiers on an identical designed data set. For this purpose, we separated 20\,\% of the data set, which corresponds to 1000 videos, as a test data set before the training. These videos were never presented to the classifiers.\\
The training of the classifiers on workflow augmented data converged with a validation loss: 0.0759 and a ACC: 0.9778 after 42 epochs of max 50 epochs for the CNN and 38 epochs of 50 epochs with a evaluation loss: 0.952 and a ACC: 0.9749 for the LSTM. Eight epochs takes for the ResNet50 approx. one day and 1 epoch per day for the LSTM on 8 NVIDIA Tesla V100 32Gb GPUs used in parallel.\\
We evaluated both classifier on the test data set. For the evaluation of the classifiers, we choose every \nth{15} image of the video due to memory issue. We evaluated the following overall metrics for multi-class imbalanced data sets \citep{Branco-2017-ID16341}: accuracy (ACC), mean accuracy (AvACC), class balanced accuracy (CBA), macro-mean recall (Rec$_\text{M}$), macro-mean precision (Prec$_\text{M}$), and macro F1-score (F1) for each network. For the calculation, the classes which do not occur in the data set are excluded.

\paragraph{\textbf{CNN}} 
The CNN classifier predicted the classes with an overall ACC $=$ 96.9\,\% and an AvACC $=$ 99.77\,\% on the workflow augmented test data set. The CBA was lower with 87.41\,\%. With a mean Prec$_\text{M}$= 96\,\%, the classifier hits the classification correctly. The Rec$_\text{M}$ was 87.5\,\% and the F1-score was 95.56\,\%.\\
\Cref{tab:bin_eval_aug_test} shows the ACC, precision (Prec), recall (Rec), specificity (Spec), and F1-score for the binary analysis of the classification. Hereby, the respective class was evaluated against all others, which were aggregated into one class. In other words, the table shows the classifier's ability to recognize the presence or absence of a specific tool class. The classifier was able to predict the classes with a probability over 99.2\,\%, with one exception: the class \textit{no tool in contact} with an ACC of 97.85\,\%. The Prec of all classifications was over 90\,\% and the Rec was over 74\,\%, with two exceptions: \textit{biomarker} with 67.9\,\% and \textit{Mendez ring\& biomarker} with 19.5\,\%. In contrast, the Spec was above 99\,\% with one exception: the class \textit{no tool in contact} with 96.7\,\%. The table also shows that the F1-score was above 0.77776, again with the exception of \textit{Mendez ring\& biomarker} of 0.3265. Additionally, we provide the complete confusion matrix for the CNN classifier on the workflow augmented test data in \Cref{tab:conf_mat_CNN_aug_test}.

\paragraph{\textbf{LSTM}} 
The LSTM classifier showed general results similar to the one of the CNN classifier. The LSTM classifier was able to predict the classes correct with an ACC = 96.6\,\% and a mean of AvACC= 99.75\,\% on the workflow augmented test data set. The CBA was lower with 88.47\,\% and a mean probability of Prec$_\text{M}$= 95.86\,\% and Rec$_\text{M}$ = 88.59\,\%, the F1-score was 92.08\,\%.\\
Further, we did a binary evaluation for the LSTM classifier. \Cref{tab:bin_eval_aug_test} shows results for the ACC, Prec, Rec, Spec, and F1-score.
The ACC scores were in the same range, from 97.5\,\% (\textit{no tool in contact}) to 100\,\% (\textit{Mendez ring\& biomarker}), as for the CNN classifier. Prec and Rec were above 86.9\,\% and above 74.8\,\%, respectively, which was lower than for CNN excluding the outlier. However, the Rec outlier for the LSTM classifier was lower at 49.88\,\% (\textit{biomarker}). The Spec was above 99.5\,\% for the LSTM with the same exception of the class \textit{no tool in contact} with 96\,\% as for CNN. The table also shows that the F1-score was above 0.8146 with the exception of \textit{biomarker}, which was 0.6514. We also provide the complete confusion matrix for the LSTM classifier in the \Cref{tab:conf_mat_LSTM_aug_test}.

The first result is: the accuracy of the two networks is similar and does not differ meaningfully from each other. Furthermore, the complexity of the CNN is high enough to be able to extract appropriate features and thereby, correctly classify the tools and tool combinations. However, from the very similar results of both networks, we cannot deduce whether the temporal component that we aimed to augment with our approach was taken into account in the classification for the LSTM classifier. The good results for both classifications, could be due to the small differences within the workflow augmented data set. Therefore, we test the classification again using original clinical videos. We also need to check whether the networks are only sensitive to the artificial videos, which may only match the original videos at first glance, but have nothing in common with real videos. For this purpose, we take the exclude video of the original training data set and the test data set to retest the networks. This was our second experiment.

\begin{table*}[ht]
\centering
\caption{Binary analysis of the classification for the workflow trained CNN and LSTM classifier on the test data set with the scores for accuracy (ACC), precision (Prec), recall (Rec), specificity (Spec), and F1-score, rounded on 4 digits. Entries with n/a could not be calculated, because they were not predicted and therefore, they are 0.}
\resizebox{0.95\textwidth}{!}{
\begin{tabular}{l||M{6ex}M{6ex}M{6ex}M{6ex}M{6ex}||M{6ex}M{6ex}M{6ex}M{6ex}M{6ex}} 
\toprule
& \multicolumn{5}{c||}{CNN} & \multicolumn{5}{c}{LSTM} \\ \cline{2-11} 
 & ACC & Prec & Rec & Spec & F1-score & ACC & Prec & Rec & Spec & F1-score \\ \hline
no tool in contact & 0.9023 & 0.8153 & 0.9854 & 0.8443 & 0.8923 & 0.9673 & 0.9426 & 0.9801 & 0.9584 & 0.9610 \\
biomarker & 0.9995 & n/a & 0 & 1 & 0 & 0.9995 & n/a & 0 & 1 & 0 \\
hydrodissection cannula & 0.9929 & 0.8210 & 0.7992 & 0.9966 & 0.8100 & 0.9973 & 0.9313 & 0.9242 & 0.9987 & 0.9278 \\
Rycroft cannula & 0.9778 & 0.9375 & 0.3870 & 0.9991 & 0.5478 & 0.9881 & 0.9109 & 0.7276 & 0.9974 & 0.8090 \\
viscoelastic cannula & 0.9871 & 0.5642 & 0.8311 & 0.9896 & 0.6721 & 0.9961 & 0.8833 & 0.8694 & 0.9981 & 0.8763 \\
cotton & 0.9999 & n/a & 0 & 1 & 0 & 0.9999 & n/a & 0 & 1 & 0 \\
capsulorhexis cystotome & 0.9818 & 0.9510 & 0.7375 & 0.9976 & 0.8308 & 0.9949 & 0.9562 & 0.9590 & 0.9972 & 0.9576 \\
Bonn forceps & 0.9997 & 0 & 0 & 1 & 0 & 0.9997 & 0 & 0 & 1 & 0 \\
capsulorhexis forceps & 0.9962 & 0.7526 & 0.4740 & 0.9991 & 0.5817 & 0.9980 & 0.9900 & 0.6429 & 1 & 0.7795 \\
Troutman forceps & 0.9997 & 0 & 0 & 0.9998 & 0 & 0.9999 & n/a & 0 & 1 & 0 \\
irrigation/aspiration handpiece & 0.9454 & 0.7827 & 0.9377 & 0.9469 & 0.8533 & 0.9814 & 0.9131 & 0.9837 & 0.9809 & 0.9471 \\
phacoemulsifier handpiece & 0.9804 & 0.8818 & 0.6056 & 0.9965 & 0.7181 & 0.9943 & 0.9295 & 0.9319 & 0.9970 & 0.9307 \\
implant injector & 0.9947 & 0.8186 & 0.8486 & 0.9970 & 0.8333 & 0.9978 & 0.9820 & 0.8761 & 0.9997 & 0.9261 \\
primary incision knife & 0.9970 & 0.8704 & 0.5697 & 0.9995 & 0.6886 & 0.9971 & 0.8268 & 0.6364 & 0.9992 & 0.7192 \\
secondary incision knife & 0.9971 & 0.7500 & 0.3645 & 0.9995 & 0.4906 & 0.9976 & 0.8421 & 0.4486 & 0.9997 & 0.5854 \\
micromanipulator & 0.9887 & 0.6115 & 0.7035 & 0.9931 & 0.6543 & 0.9962 & 0.8997 & 0.8447 & 0.9985 & 0.8714 \\
suture needle & 1 & n/a & n/a & 1 & n/a & 1 & n/a & n/a & 1 & n/a \\
Mendez ring & 0.9986 & n/a & 0 & 1 & 0 & 0.9987 & 1 & 0.0263 & 1 & 0.0513 \\
\begin{tabular}[c]{@{}l@{}}Mendez ring \&\\ biomarker\end{tabular} & 0.9999 & n/a & 0 & 1 & 0 & 0.9999 & n/a & 0 & 1 & 0 \\
\begin{tabular}[c]{@{}l@{}}Bonn forceps \& \\ secondary incision knife\end{tabular} & 0.9988 & 0.7191 & 0.8767 & 0.9991 & 0.7901 & 0.9988 & 0.7941 & 0.7397 & 0.9995 & 0.7660 \\
\begin{tabular}[c]{@{}l@{}}primary incision knife \& \\ Bonn forceps\end{tabular} & 0.9980 & 0.4464 & 0.5000 & 0.9989 & 0.4717 & 0.9985 & 0.6000 & 0.4800 & 0.9994 & 0.5333 \\
\begin{tabular}[c]{@{}l@{}}capsulorhexis cystotome \& \\ Bonn forceps\end{tabular} & 0.9996 & 0 & 0 & 0.9999 & 0 & 0.9997 & 0 & 0 & 1 & 0 \\
\begin{tabular}[c]{@{}l@{}}phacoemulsifier handpiece \& \\ Bonn forceps\end{tabular} & 0.9989 & 1 & 0.1622 & 1 & 0.2791 & 0.9987 & 1 & 0.0270 & 1 & 0.0526 \\
\begin{tabular}[c]{@{}l@{}}phacoemulsifier handpiece \& \\ micromanipulator\end{tabular} & 0.9414 & 0.9430 & 0.6443 & 0.9932 & 0.7655 & 0.9888 & 0.9512 & 0.9749 & 0.9913 & 0.9629 \\
\begin{tabular}[c]{@{}l@{}}irrigation/aspiration handpiece \& \\ micromanipulator\end{tabular} & 0.9672 & 0.8352 & 0.3475 & 0.9967 & 0.4908 & 0.9818 & 0.9320 & 0.6478 & 0.9977 & 0.7643 \\
\begin{tabular}[c]{@{}l@{}}hydrodissection cannula \& \\ micromanipulator\end{tabular} & 1 & 0 & n/a & 1 & 0 & 1 & n/a & n/a & 1 & n/a \\
\begin{tabular}[c]{@{}l@{}}Troutman forceps \& \\ suture needle\end{tabular} & 1 & n/a & n/a & 1 & n/a & 1 & n/a & n/a & 1 & n/a \\
\bottomrule
\end{tabular}
}
\label{tab:bin_eval_real_test}
\end{table*}

\begin{table*}[ht]
\centering
\caption{Binary analysis of the classification for the split trained CNN and LSTM classifier on the test data set with the scores for accuracy (ACC), precision (Prec), recall (Rec), specificity (Spec), and F1-score, rounded on 4 digits. Entries with n/a could not be calculated, because they were not predicted and therefore, they are 0.}
\resizebox{0.95\textwidth}{!}{
\begin{tabular}{l||M{6ex}M{6ex}M{6ex}M{6ex}M{6ex}||M{6ex}M{6ex}M{6ex}M{6ex}M{6ex}} 
\toprule
& \multicolumn{5}{c||}{CNN} & \multicolumn{5}{c}{LSTM} \\ \cline{2-11} 
 & ACC & Prec & Rec & Spec & F1-score & ACC & Prec & Rec & Spec & F1-score \\ \hline
no tool in contact & 0.9050 & 0.8335 & 0.9607 & 0.8662 & 0.8335 & 0.9619 & 0.9391 & 0.9700 & 0.9562 & 0.9543 \\
biomarker & 0.9995 & n/a & 0 & 1 & n/a & 0.9995 & n/a & 0 & 1 & 0 \\
hydrodissection cannula & 0.9791 & 0.4522 & 0.4924 & 0.9885 & 0.4522 & 0.9938 & 0.8447 & 0.8239 & 0.9971 & 0.8341 \\
Rycroft cannula & 0.9742 & 0.6611 & 0.5273 & 0.9903 & 0.6611 & 0.9808 & 0.6907 & 0.8091 & 0.9870 & 0.7452 \\
viscoelastic cannula & 0.9762 & 0.3744 & 0.7387 & 0.9800 & 0.3744 & 0.9915 & 0.7353 & 0.7320 & 0.9957 & 0.7336 \\
cotton & 0.9999 & n/a & 0 & 1 & n/a & 0.9999 & n/a & 0 & 1 & 0 \\
capsulorhexis cystotome & 0.9657 & 0.8542 & 0.5220 & 0.9943 & 0.8542 & 0.9886 & 0.9145 & 0.8955 & 0.9946 & 0.9049 \\
Bonn forceps & 0.9997 & n/a & 0 & 1 & n/a & 0.9997 & n/a & 0 & 1 & 0 \\
capsulorhexis forceps & 0.9941 & 0.3830 & 0.1169 & 0.9990 & 0.3830 & 0.9954 & 0.6582 & 0.3377 & 0.9990 & 0.4464 \\
Troutman forceps & 0.9994 & 0 & 0 & 0.9995 & 0 & 0.9999 & n/a & 0 & 1 & 0 \\
irrigation/aspiration handpiece & 0.9138 & 0.7222 & 0.7978 & 0.9374 & 0.7222 & 0.9819 & 0.9318 & 0.9636 & 0.9856 & 0.9474 \\
phacoemulsifier handpiece & 0.9625 & 0.5526 & 0.4677 & 0.9838 & 0.5526 & 0.9901 & 0.9274 & 0.8246 & 0.9972 & 0.8730 \\
implant injector & 0.9917 & 0.7164 & 0.7821 & 0.9951 & 0.7164 & 0.9959 & 0.8812 & 0.8509 & 0.9982 & 0.8658 \\
primary incision knife & 0.9956 & 0.7590 & 0.3818 & 0.9993 & 0.7590 & 0.9952 & 0.5812 & 0.6727 & 0.9971 & 0.6236 \\
secondary incision knife & 0.9969 & 0.7442 & 0.2991 & 0.9996 & 0.7442 & 0.9967 & 0.5660 & 0.5607 & 0.9983 & 0.5634 \\
micromanipulator & 0.9811 & 0.4056 & 0.5106 & 0.9884 & 0.4056 & 0.9881 & 0.6186 & 0.5647 & 0.9946 & 0.5904 \\
suture needle & 1 & n/a & n/a & 1 & n/a & 1 & n/a & n/a & 1 & n/a \\
Mendez ring & 0.9986 & 0 & 0 & 1 & 0 & 0.9986 & n/a & 0 & 1 & 0 \\
\begin{tabular}[c]{@{}l@{}}Mendez ring \&\\ biomarker\end{tabular} & 0.9999 & n/a & 0 & 1 & n/a & 0.9999 & n/a & 0 & 1 & 0 \\
\begin{tabular}[c]{@{}l@{}}Bonn forceps \& \\ secondary incision knife\end{tabular} & 0.9980 & 0.6545 & 0.4932 & 0.9993 & 0.6545 & 0.9980 & 1 & 0.2192 & 1 & 0.3596 \\
\begin{tabular}[c]{@{}l@{}}primary incision knife \& \\ Bonn forceps\end{tabular} & 0.9977 & 0.3684 & 0.4200 & 0.9987 & 0.3684 & 0.9982 & 0 & 0 & 1 & 0 \\
\begin{tabular}[c]{@{}l@{}}capsulorhexis cystotome \& \\ Bonn forceps\end{tabular} & 0.9996 & 0 & 0 & 0.9999 & 0 & 0.9997 & n/a & 0 & 1 & 0 \\
\begin{tabular}[c]{@{}l@{}}phacoemulsifier handpiece \& \\ Bonn forceps\end{tabular} & 0.9980 & 0.1538 & 0.1081 & 0.9992 & 0.1538 & 0.9987 & n/a & 0 & 1 & 0 \\
\begin{tabular}[c]{@{}l@{}}phacoemulsifier handpiece \& \\ micromanipulator\end{tabular} & 0.9009 & 0.7203 & 0.5433 & 0.9632 & 0.7203 & 0.9845 & 0.9515 & 0.9439 & 0.9916 & 0.9477 \\
\begin{tabular}[c]{@{}l@{}}irrigation/aspiration handpiece \& \\ micromanipulator\end{tabular} & 0.9636 & 0.6649 & 0.4019 & 0.9903 & 0.6649 & 0.9809 & 0.8367 & 0.7226 & 0.9933 & 0.7755 \\
\begin{tabular}[c]{@{}l@{}}hydrodissection cannula \& \\ micromanipulator\end{tabular} & 0.9997 & 0 & n/a & 0.9997 & 0 & 1 & n/a & n/a & 1 & n/a \\
\begin{tabular}[c]{@{}l@{}}Troutman forceps \& \\ suture needle\end{tabular} & 1 & n/a & n/a & 1 & n/a & 1 & n/a & n/a & 1 & n/a \\ \bottomrule
\end{tabular}%
}
\label{tab:bin_eval_split_real_test}
\end{table*}

\subsection{Classification of the networks on real videos}
\label{subsec:classificaton_real_data}
\cite{dataset_cataract} provided during our study the test data with the classification in addition to the training data set. Therefore, we had the choice to test our models on real cataract surgery videos from the test data set and not only on video 4 from the training data, as mention in the beginning of the manuscript. First, we must determine which tools and tool combinations are present in the test data, due to the fact that not all were present in our training data set. This is important because classes that were not included in the training data can strongly negatively bias the results. Especially if the temporal sequence is taken into account in addition to the spatial information in the individual frames, as it is for the LSTM model. After evaluating the containing tool classes, we have to exclude 8 of the 25 videos. The real test data set now contains only 17 in addition to video 4 of the training data set. \Cref{tab:tooldistribution} show for each class the distribution of the test data set. However, the test data set does not include all instrument classes. The following classes was not observed in the selected videos: \textit{suture needle}, \textit{hydrodissection cannula \& micromanipulator}, and \textit{Troutman forceps \& suture needle}.

\subsubsection{Trained on the workflow augmented data set}
In the following, we present results of the two classifiers that are trained on the workflow augmented data set and retested on the real video.

\paragraph{\textbf{CNN}} On the real test data set, the CNN classifier was able to predict the classes with an ACC of 82.12\,\% and an AvACC of 98.68\,\%. The CBA was 39.4\,\% excluding the classes were row and column are zero e.g. \textit{suture needle}, \textit{Troutman forceps \& suture needle}, seen \Cref{APP:tab:conf_mat_CNN_test}. The mean Prec$_\text{M}$ was 64.29\,\%. The mean Rec$_\text{M}$ was 44.89\,\% and the F1-score was 52.86\,\%. The results of the CNN classifier were markedly worse as for the workflow augmented test data set. The ACC was 12 percentage points (pp) lower on the real data set than on the workflow augmented test data set. All other scores were also lower.\\
The binary analysis in \Cref{{tab:bin_eval_real_test}} showed, that the classes were detected with an ACC of over 94.14\,\%. The class \textit{no tool in contact}, which occurred most in the training data, had the lowest ACC with 90.23\,\%. However, the high ACC was not achieved by the correctly recognized classes, it was achieved by the correct classification of the non-classes, which can be seen in the Prec and Rec scores. The Prec can partly not be calculated, e.g.: \textit{biomarker}, \textit{Mendez ring}, or \textit{Mendez ring \& biomarker}, because these classes were not predicted at all. The Prec score for this classes was n/a. For the classes \textit{Bonn forceps}, \textit{Troutman forceps}, \textit{capsulorhexis cystotome \& Bonn forceps} and \textit{hydrodissection cannula \& micromanipulator} the Prec was zero due to less then 5 false-positive predictions per class. For the other classes, the Prec was in the range from 44.64\,\% to 100\,\%. The Rec was in the range 0\,\% to 98.54\,\%. Hereby, 11 classes were 0\,\%, because they were never correctly classified, which can be observed in the confusion matrix for the CNN classifier in \Cref{APP:tab:conf_mat_CNN_test}. Mostly, the CNN classified these classes as the class \textit{no tool in contact}. However, the Spec for the class \textit{no tool in contact} with 84.43\,\%, also the lowest score compared to the other classes. Here, the Spec was above 94\,\%.

\paragraph{\textbf{LSTM}}
The classification results of the LSTM were also worst on the videos of the real test data set than on the workflow augmented test data set. Nevertheless, they were much better than the results of the CNN. The ACC was 93.49\,\% that was over 11 pp higher as the ACC of the CNN. The AvACC was 99.52\,\% and the CBA was 52.43\,\% excluded classes, for which the rows and columns were zero. The Prec$_\text{M}$ was 81.42\,\%, which was also better than the CNN with over 17 pp. The mean Rec$_\text{M}$ was 53\,\% and the F1-score was 64.2\,\%. In the binary analysis of the LSTM, which is show in \Cref{tab:bin_eval_real_test}, the LSTM perform generally better than the CNN. The ACC was higher than 96.73\,\% for every class. However, the class \textit{no tool in contact} also had the worst recognition rate. Furthermore, the Prec was only in two classes 0\,\% and in four classes n/a besides the classes that are not included in the test data. For the other classes the Prec was at least above 60\,\% (\textit{primary incision knife \& Bonn forceps}), but in the most cases above 82\,\%. The high ACC of the LSTM classifier was not based on the high true-negative rates like for the CNN classifier, but on higher true-positive rates. For each classification, the score was either higher or at least equivalent, except for two classes: \textit{primary incision knife \& Bonn forceps} and \textit{phacoemulsifier handpiece \& Bonn forceps}, which had lower scores. \Cref{APP:tab:conf_mat_LSTM_test} shows that if a frame was misclassified, it was mostly placed in the class \textit{no tool in contact} or in a neighboring class, e.g.: the class \textit{secondary incision knife} was classified as \textit{Bonn forceps \& secondary incision knife} in 12 cases. More examples can be found in the table. In contrast to \Cref{APP:tab:conf_mat_CNN_test}, the misclassification scatter of the LSTM was lower than that of the CNN.

Another result is: the LSTM must had taken into account the temporal information during the training, otherwise a difference between CNN and LSTM network would not be observable. This raises another question: can the result also be achieved with a less complicated method such as split data augmentation? We answer this question with our third experiment. 

\subsubsection{Trained on the split data set}
\label{subsubsec:res:triained_split}
To benchmark the results of our approach, we retrained and evaluated both network CNN and LSTM as described in \Cref{subsec:training_strategy}. We used the split data set, which were augmented based on the video splitting approach as descried in \Cref{subsubsec:split_augmentation}.\\ 
We applied the same termination criterion in training as before on the workflow augmented data set. The loss for the CNN converged after 16 of 50 epochs with a ACC of 98.54\,\% and a loss of 0.1092 for the training data set and an ACC of 55.18\,\% for the evaluation data set. The loss of the LSTM net converged after 19 of 50 epochs with an ACC of 98.61\,\% and a loss of 0.0823 for training data set and an evaluation ACC of 80.46\,\%. One epoch takes for the CNN approx. 1 hour and for the LSTM approx. 1.5 hour on 8 NVIDIA Tesla V100 32Gb GPUs used in parallel, which was significantly faster than for other data set. Afterwards, we evaluated the networks using the same parameters on the real test data as descried previously. 

\paragraph{\textbf{CNN}}
The CNN achieved an ACC of 75.52\,\% and an AvACC of 98.11\,\%. The CBA was 30.94\,\%. Thereby, the ACC and the CBA of the split trained networks were lower with about 8 pp as the scores of the CNN trained on the workflow augmented data. The AvACC was also lower, but in the same order of magnitude.
The scores for Prec$_M$ with 47.72\,\%, the Rec$_M$ with 35.68\,\% and the F1-score with 40.83\,\% show that the classifier generally performs worse than the workflow trained CNN.\\
The binary analysis, shown in \Cref{tab:bin_eval_split_real_test}, shows that the ACC for the different classes was in the same order of magnitude as the ACC of the workflow trained CNN, but the Prec does not achieve more than 85.42\,\%. In comparison, for fewer classes the Prec was 0\,\%, but for example, for the class \textit{phacoemulsifier handpiece \& Bonn forceps} the Prec was 85 pp lower as for the other CNN. The median for the Prec was 66.49\,\%. The scores for Rec was generally lower than the workflow trained CNN, but there are two exceptions for class \textit{Rycroft cannula} and \textit{irrigation/aspiration handpiece \& micromanipulator}: the Rec increased to 52.73\,\% (previously 38.7\,\%) and 40.19\,\% (previously 34.75\,\%), respectively.
Since the scores for Prec and Rec were lower, the values for the F1-score were also lower for the individual classes with the exception of the two classes mentioned above. The Spec was similar to those of the workflow trained CNN. The confusion matrix \Cref{tab:conf_mat_CNN_split_test} seen in \Cref{sec:App_conf} also shows more entries that are not 0 as int the \Cref{tab:conf_mat_CNN_aug_test}.

\paragraph{\textbf{LSTM}}
The LSTM classifier achieved an ACC of 90.87\,\% on the real test data and an AvACC of 99.32\,\%. The CBA was 44.24\,\%. Thereby, the three scores of the split trained networks were lower as the scores of the LSTM trained on the workflow augmented data set. Also the scores for mean Prec$_M$ with 75.48\,\%, the mean Rec$_M$ with 45.38\,\% and the F1-score with 56.68\,\% showed that the classifier generally performs worse than the workflow trained LSTM. The scores was $\approx 6$ pp greater for Prec$_M$ and Rec$_M$, and for the F1-score it was $\approx 8$ pp higher.
The binary analysis of the split trained LSTM (\Cref{tab:bin_eval_split_real_test}) yielded similar scores for the ACC of the classes as the workflow trained. The scores of Prec were mostly worse, only five classes reached scores above 90\,\%, compared to ten before. In addition, the Prec could not be calculated for 8 classes in addition to the three which could not observed in the real test data. They were neither true nor false-positive predicted by the network, see \Cref{tab:bin_eval_split_real_test}. Similar results could also be obtained for the Rec scores. Hereby, the scores were 0\,\% for eight classes and three classes were above 90\,\% instead of six. This trend was also visible for the F1-score.\\
Furthermore, in the confusion matrix \Cref{tab:conf_mat_LSTM_split_test} in \Cref{sec:App_conf} it can be observed that more entries were not 0 as the comparison confusion matrix \Cref{tab:conf_mat_LSTM_aug_test}. 

Finally, we conclude that the neural networks that were trained on the workflow augmented data set was the best. Furthermore, the artificially generated video for the neural network is indistinguishable from the real video, or would not have a negative effect on the recognition performance of the classifiers. 

\section{Discussion}
\label{sec:discussion}
One goal of our workflow augmentation approach was to create new artificial videos with a comparable duration as the original videos.
This goal could be achieved by choosing the initial transition probabilities to be uniformly distributed. This resulted not only in videos whose duration is more balanced, but also the alternation frequency of classes is higher in videos while keeping a very similar length of the individual segment, see \Cref{tab:Comparison_workflow_org}. The median segment length is slightly shorter with 121\,frames compared to 126\,frames of the original data set. This can be explained by the fact that the sub-frames variation of temporal augmentation has a median score of 67.5\,frames instead of 64\,frames. However, this should not lead to any noticeable effect on the networks.

Furthermore, an aim of the approach was that the prevalence of the individual classes within the data set would be more balanced. So that recognition performance of classes that were underrepresented in the original dataset would be improved. The results in \Cref{subsec:res:Workflow_augmentation_data set} showed that our workflow augmented data set is more balanced as the original data set, although not for all classes. This can be obtained from \Cref{tab:tooldistribution}. For example, the percentage of the class \textit{Troutman forceps} was increased by a factor of 8.8, but the presence of the previously dominated class \textit{no tool in contact} was also increased from 46.30\,\% to 50.06\,\%. However, the relatively underrepresented classes now appear more frequently, but the distribution of the classes is still not uniform. The goal of equal class distribution might be too ambitious and could not be achieved prospectively with the chosen transition probabilities. The underlying goal of better detection of underrepresented classes could perhaps still be achieved by our approach.
Therefore, we looked at the different experiments. The first experiment in \Cref{subsec:classification_workflow_test} shows that the selected ResNet50 is able to separate and classify the individual classes.
Unfortunately, the addition of the LSTM layer did not yield any significant improvements. This could be due to the fact that the CNN had already generalized very well due to the large number of training data images, averaging 43.5 million.

Whether underrepresented classes are better detected by training the network with a data set, augmented by our approach, cannot be answered by this experiment, because the ACC of the binary analysis for all classes is above 99\,\%, except for the class no tool in contact. Also, the scores for Prec and Rec are not unambiguous. These values are also valid for the LSTM. However, the results in \Cref{subsec:classificaton_real_data} of the second experiment show considerable differences from the results of the first experiment. From the results in \Cref{tab:bin_eval_real_test}, we conclude that the additional LSTM layer leads to an improvement. All values of the ACC for the classification of LSTM network are at least as high or higher than the one for the CNN. The F1-score is also higher for the LSTM network, but there are 2 exceptions. For the class \textit{Bonn forceps \& secondary incision knife} and the class \textit{phacoemulsifier handpiece \& Bonn forceps} the F1-core is lower for the LSTM as for the CNN network. This is due to the worse score for the Rec. The class \textit{phacoemulsifier handpiece \& Bonn forceps} is also an example of an underrepresented class with 8.3\,\%. It is the second least common class next to the \textit{Biomaker} class, which is the third most common, and the \textit{Mendez ring \& biomarker} the rarest class. In contrast to the class \textit{phacoemulsifier handpiece \& Bonn forceps}, the Prec for \textit{Biomaker} and \textit{Mendez ring \& biomarker} could not be calculated, because these classes were not predicted by the LSTM network, see \Cref{APP:tab:conf_mat_LSTM_test}. Furthermore, classes that occur more frequently in the training data but very rarely in the real test data, such as \textit{cotton} or \textit{Troutman forceps}, were also not predicted by the network. These classes are mostly detected exclusively at \textit{not tool in contact}, see \Cref{APP:tab:conf_mat_LSTM_test}. This leads to the assumption that either the timing of the tool in tissue contact is not 100\,\% correct, or that the network has learned to identify every image that is not recognized as a different class as \textit{not tool in contact}. Furthermore, it is surprising that the class \textit{no tool in contact}, which was most common in the training data, in the test data of the workflow-enhanced dataset as well as in the real test data, performed worst for the ACC in the binary evaluation. This also partially supports our assumption of timing uncertainty.

However, in comparison with \cite{Hajj-2017-ID16202}, both networks trained on the workflow augmented data set, created by our approach and tested on the real test data, are performing better. The mean precision of our CNN network is with 64.29\,\% and 81.42\,\% for the LSTM network higher than the pure CNN and the combined CNN and RNN network from \cite{Hajj-2017-ID16202}, which had a mean precision between 52.93\,\% and 60.86\,\% for the CNN and 79.80\,\% for the combined network. Furthermore, in contrast to Al Hajj~et~al, we did not perform any hyper parameter tuning. Nevertheless, the results are not directly comparable because not all instruments were included in our workflow graph. For a better comparison we consequently use the same augmentation method as \cite{Hajj-2017-ID16202} with the same conditions as we use to create the workflow augmented data set, e.g. types of instruments, spatial augmentation parameter, etc. and retrain both networks on this data. This was other third experiment.

Hereby, we access also a better recognition performance for the networks that were trained with the data set that was augmented by our approach. The ACC for the CNN is with 82.12\,\% higher then 75.52\,\% from the CNN that are trained on the split data set. Also Prec$_M$ with 64.29\,\% and Rec$_M$ with 44.89\,\% are compared to 47.72\,\% and 35.68\,\%, respectively, are higher for the CNN trained on the workflow augmented data. This leads to the assumption that the split data set is in general too small for a suitable training of a neural network like the ResNet50. Since the networks were only trained on spatially augmented images and do not contain any new additional augmented information. For the LSTM network, we also obtained remarkably better results for the model that are trained using our new approach. Hereby the ACC is 93.49\,\% compared to 90.87\,\% for the split trained, with a Prec$_M$ of 81.42 and Rec$_M$ of 53\,\% and 75.48\,\% and 45.38\,\%, respectively, for the state of the art augmentation method. In contrast to the CNN networks, we were able to augment not only the image information but also the temporal sequence, otherwise the results of CNN and LSTM would not be so different. Finally, we state that the classifier trained on data that are augmented with our approach provides substantially better results in terms of tool recognition compared to the state of the art method.

However, we need to address some limitations. Not all training videos were taken into account when creating the workflow diagrams. Therefore, direct comparisons cannot be made without constraints on the other publication. Also, no explicit thresholds are used for classification, instead the one-hot classifier was used, which means that the threshold for a class can be different for each image, but the score for the class must be the highest. Furthermore, a multi-class approach was used instead of a multi-label approach, but this should not result in any overall differences. Also, a weighted cost function could result in further improvements, as the data set is more balanced than the original data set but the balancing of the data set is still not enough. 

\section{Conclusion}
\label{sec:conclusion}
In this paper, we introduce a novel approach for augmenting videos in the field video-based event recognition. This methodology allows creating new artificial videos that appear to the neural network as an original recorded video. Furthermore, it is possible to create sequences of the same duration, with variation of speed, and to balance the classifications in data sets a posteriori.\\
Furthermore, the proposed approach has two novelties. The first novelty is that a combined CNN and RNN network can be trained end-to-end, since sufficient data can now be generated for training. The second novelty is the methodology of augmentation: by combining meta knowledge with the workflow, spatial and temporal augmentation, it is now possible to balance and completely augment temporal sequences such as videos of a data set. The proposed methodology is general and applicable outside the scope of cataract surgery video analysis. \\
We have shown that our approach is capable of extending and balancing small data sets, using only information that is contained in the data set. Furthermore, our method is not limited to only this information. We designed a benchmark experiment in which we trained two exactly identical networks with two different data sets. One with a data set created by our new proposed approach and one with a data set created by an established approach. For both data sets, we used exactly the same original data and parameters. \\
Compared to current approaches, the network that are trained with our workflow augmented data achieve a better classification accuracy than the comparison network. Based on our preliminary results, we believe that our proposed approach has a high potential to improve the video classification and recognition not only in the medical field. 
In future work, the approach will to be validated and its potential in other fields like gait analysis or further applications will to be shown. In addition, the workflow model could be extended and refined by other information, e.g. anatomical data or special patient data, because physicians often use data on patient history, age, demographics for the decision-making process.\\

\noindent Finally, we were able to show that it is advantageous to build a workflow model from annotated data. The synthetic data created with the workflow augmentation lead to better classification result by a neural networks.

\section*{Acknowledgments}
The authors would like to thank Steffen Schuler valuable discussions.\\
The authors would also like to thank~\cite{dataset_cataract} for sharing the cataract data set and the support by the state of Baden-Württemberg through bwHPC.

\appendix
\section{Confusion matrix of different classifier}
\label{sec:App_conf}

\begin{table*}[t]
\Huge
\centering
\caption{Confusion matrix rounded to 4 digits of the workflow augmented test data for the CNN classifier.}
\resizebox{\textwidth}{!}{%
\def\arraystretch{1.5}
\begin{tabular}{|c||c|c|c|c|c|c|c|c|c|c|c|c|c|c|c|c|c|c|c|c|c|c|c|c|c|c|c||c|} 
\hline
\diagbox[dir=NW, width = 7.5cm] {\rule{0cm}{6.5cm}{\textbf{\Huge{Label}}}}{\raisebox{-0.5ex}{\textbf{\Huge{Prediction}}}} & \rot{\parbox[*][7ex][c]{6cm}{no tool in \\contact} } & \rot{\parbox[*][7ex][c]{6cm}{biomarker}} & \rot{\parbox[*][7ex][c]{6cm}{hydrodissection \\cannula}} & \rot{\parbox[*][7ex][c]{6cm}{Rycroft cannula}} & \rot{\parbox[*][7ex][c]{6cm}{viscoelastic \\cannula}} & \rot{\parbox[*][7ex][c]{6cm}{cotton}} & \rot{\parbox[*][7ex][c]{6cm}{capsulorhexis \\cystotome}} & \rot{\parbox[*][7ex][c]{6cm}{Bonn forceps}} & \rot{\parbox[*][7ex][c]{6cm}{capsulorhexis \\forceps}} & \rot{\parbox[*][7ex][c]{6cm}{Troutman \\forceps}} & \rot{\parbox[*][7ex][c]{6cm}{irrigation/aspiration handpiece}} & \rot{\parbox[*][7ex][c]{6cm}{phacoemulsifier \\handpiece}} & \rot{\parbox[*][7ex][c]{6cm}{implant \\injector}} & \rot{\parbox[*][7ex][c]{6cm}{primary \\incision knife}} & \rot{\parbox[*][7ex][c]{6cm}{secondary \\incision knife}} & \rot{\parbox[*][7ex][c]{6cm}{micromanipulator}} & \rot{\parbox[*][7ex][c]{6cm}{suture needle}} & \rot{\parbox[*][7ex][c]{6cm}{Mendez ring} } & \rot{\parbox[*][7ex][c]{6cm}{Mendez ring \& \\biomarker}} & \rot{\parbox[*][7ex][c]{6cm}{Bonn forceps \& \\ secondary \\incision knife}} & \rot{\parbox[*][7ex][c]{6cm}{primary \\incision knife \& \\Bonn forceps}} & \rot{\parbox[*][7ex][c]{6cm}{capsulorhexis \\cystotome \&\\ Bonn forceps}} & \rot{\parbox[*][7ex][c]{6cm}{phacoemulsifier \\handpiece \&\\ Bonn forceps}} & \rot{\parbox[*][7ex][c]{6cm}{phacoemulsifier \\handpiece \&\\ micromanipulator}} & \rot{\parbox[*][7ex][c]{6cm}{irrigation/aspiration \\handpiece \& \\ micromanipulator}} & \rot{\parbox[*][7ex][c]{6cm}{hydrodissection \\cannula \& \\ micromanipulator}} 
& \rot{\parbox[*][7ex][c]{6cm}{Troutman \\forceps \& \\ suture needle}} & \rot{\parbox[*][7ex][c]{6cm}{Recall}} \\
\hline
\end{tabular}%
}
\label{tab:conf_mat_CNN_aug_test}
\end{table*}

\begin{table*}[t]
\Huge
\def\arraystretch{1.5}
\centering
\caption{Confusion matrix rounded to 4 digits of the workflow augmented test data for the LSTM classifier.}
\resizebox{\textwidth}{!}{%
\begin{tabular}{|c||c|c|c|c|c|c|c|c|c|c|c|c|c|c|c|c|c|c|c|c|c|c|c|c|c|c|c||c|} 
\hline
\diagbox[dir=NW, width = 7.5cm] {\rule{0cm}{6.5cm}{\textbf{\Huge{Label}}}}{\raisebox{-0.5ex}{\textbf{\Huge{Prediction}}}} & \rot{\parbox[*][7ex][c]{6cm}{no tool in \\contact} } & \rot{\parbox[*][7ex][c]{6cm}{biomarker}} & \rot{\parbox[*][7ex][c]{6cm}{hydrodissection \\cannula}} & \rot{\parbox[*][7ex][c]{6cm}{Rycroft cannula}} & \rot{\parbox[*][7ex][c]{6cm}{viscoelastic \\cannula}} & \rot{\parbox[*][7ex][c]{6cm}{cotton}} & \rot{\parbox[*][7ex][c]{6cm}{capsulorhexis \\cystotome}} & \rot{\parbox[*][7ex][c]{6cm}{Bonn forceps}} & \rot{\parbox[*][7ex][c]{6cm}{capsulorhexis \\forceps}} & \rot{\parbox[*][7ex][c]{6cm}{Troutman \\forceps}} & \rot{\parbox[*][7ex][c]{6cm}{irrigation/aspiration handpiece}} & \rot{\parbox[*][7ex][c]{6cm}{phacoemulsifier \\handpiece}} & \rot{\parbox[*][7ex][c]{6cm}{implant \\injector}} & \rot{\parbox[*][7ex][c]{6cm}{primary \\incision knife}} & \rot{\parbox[*][7ex][c]{6cm}{secondary \\incision knife}} & \rot{\parbox[*][7ex][c]{6cm}{micromanipulator}} & \rot{\parbox[*][7ex][c]{6cm}{suture needle}} & \rot{\parbox[*][7ex][c]{6cm}{Mendez ring} } & \rot{\parbox[*][7ex][c]{6cm}{Mendez ring \& \\biomarker}} & \rot{\parbox[*][7ex][c]{6cm}{Bonn forceps \& \\ secondary \\incision knife}} & \rot{\parbox[*][7ex][c]{6cm}{primary \\incision knife \& \\Bonn forceps}} & \rot{\parbox[*][7ex][c]{6cm}{capsulorhexis \\cystotome \&\\ Bonn forceps}} & \rot{\parbox[*][7ex][c]{6cm}{phacoemulsifier \\handpiece \&\\ Bonn forceps}} & \rot{\parbox[*][7ex][c]{6cm}{phacoemulsifier \\handpiece \&\\ micromanipulator}} & \rot{\parbox[*][7ex][c]{6cm}{irrigation/aspiration \\handpiece \& \\ micromanipulator}} & \rot{\parbox[*][7ex][c]{6cm}{hydrodissection \\cannula \& \\ micromanipulator}} 
& \rot{\parbox[*][7ex][c]{6cm}{Troutman \\forceps \& \\ suture needle}} & \rot{\parbox[*][7ex][c]{6cm}{Recall}} \\
\hline
\end{tabular}%
}
\label{tab:conf_mat_LSTM_aug_test}
\end{table*}
\begin{table*}[t]
\Huge
\def\arraystretch{1.5}
\centering
\caption{Confusion matrix rounded to 4 digits of the real test data set for the workflow augmented trained CNN classifier.}
\resizebox{\textwidth}{!}{%
\begin{tabular}{|c||c|c|c|c|c|c|c|c|c|c|c|c|c|c|c|c|c|c|c|c|c|c|c|c|c|c|c||c|} 
\hline
\diagbox[dir=NW, width = 7.5cm] {\rule{0cm}{6.5cm}{\textbf{\Huge{Label}}}}{\raisebox{-0.5ex}{\textbf{\Huge{Prediction}}}} & \rot{\parbox[*][7ex][c]{6cm}{no tool in \\contact} } & \rot{\parbox[*][7ex][c]{6cm}{biomarker}} & \rot{\parbox[*][7ex][c]{6cm}{hydrodissection \\cannula}} & \rot{\parbox[*][7ex][c]{6cm}{Rycroft cannula}} & \rot{\parbox[*][7ex][c]{6cm}{viscoelastic \\cannula}} & \rot{\parbox[*][7ex][c]{6cm}{cotton}} & \rot{\parbox[*][7ex][c]{6cm}{capsulorhexis \\cystotome}} & \rot{\parbox[*][7ex][c]{6cm}{Bonn forceps}} & \rot{\parbox[*][7ex][c]{6cm}{capsulorhexis \\forceps}} & \rot{\parbox[*][7ex][c]{6cm}{Troutman \\forceps}} & \rot{\parbox[*][7ex][c]{6cm}{irrigation/aspiration handpiece}} & \rot{\parbox[*][7ex][c]{6cm}{phacoemulsifier \\handpiece}} & \rot{\parbox[*][7ex][c]{6cm}{implant \\injector}} & \rot{\parbox[*][7ex][c]{6cm}{primary \\incision knife}} & \rot{\parbox[*][7ex][c]{6cm}{secondary \\incision knife}} & \rot{\parbox[*][7ex][c]{6cm}{micromanipulator}} & \rot{\parbox[*][7ex][c]{6cm}{suture needle}} & \rot{\parbox[*][7ex][c]{6cm}{Mendez ring} } & \rot{\parbox[*][7ex][c]{6cm}{Mendez ring \& \\biomarker}} & \rot{\parbox[*][7ex][c]{6cm}{Bonn forceps \& \\ secondary \\incision knife}} & \rot{\parbox[*][7ex][c]{6cm}{primary \\incision knife \& \\Bonn forceps}} & \rot{\parbox[*][7ex][c]{6cm}{capsulorhexis \\cystotome \&\\ Bonn forceps}} & \rot{\parbox[*][7ex][c]{6cm}{phacoemulsifier \\handpiece \&\\ Bonn forceps}} & \rot{\parbox[*][7ex][c]{6cm}{phacoemulsifier \\handpiece \&\\ micromanipulator}} & \rot{\parbox[*][7ex][c]{6cm}{irrigation/aspiration \\handpiece \& \\ micromanipulator}} & \rot{\parbox[*][7ex][c]{6cm}{hydrodissection \\cannula \& \\ micromanipulator}} 
& \rot{\parbox[*][7ex][c]{6cm}{Troutman \\forceps \& \\ suture needle}} & \rot{\parbox[*][7ex][c]{6cm}{Recall}} \\
\hline
\end{tabular}
}
\label{APP:tab:conf_mat_CNN_test}
\end{table*}

\begin{table*}[t]
\Huge
\def\arraystretch{1.5}
\centering
\caption{Confusion matrix rounded to 4 digits of the real test data set for the workflow augmented trained LSTM classifier.}
\resizebox{\textwidth}{!}{%
\begin{tabular}{|c||c|c|c|c|c|c|c|c|c|c|c|c|c|c|c|c|c|c|c|c|c|c|c|c|c|c|c||c|}
\hline
\diagbox[dir=NW, width = 7.5cm] {\rule{0cm}{6.5cm}{\textbf{\Huge{Label}}}}{\raisebox{-0.5ex}{\textbf{\Huge{Prediction}}}} & \rot{\parbox[*][7ex][c]{6cm}{no tool in \\contact} } & \rot{\parbox[*][7ex][c]{6cm}{biomarker}} & \rot{\parbox[*][7ex][c]{6cm}{hydrodissection \\cannula}} & \rot{\parbox[*][7ex][c]{6cm}{Rycroft cannula}} & \rot{\parbox[*][7ex][c]{6cm}{viscoelastic \\cannula}} & \rot{\parbox[*][7ex][c]{6cm}{cotton}} & \rot{\parbox[*][7ex][c]{6cm}{capsulorhexis \\cystotome}} & \rot{\parbox[*][7ex][c]{6cm}{Bonn forceps}} & \rot{\parbox[*][7ex][c]{6cm}{capsulorhexis \\forceps}} & \rot{\parbox[*][7ex][c]{6cm}{Troutman \\forceps}} & \rot{\parbox[*][7ex][c]{6cm}{irrigation/aspiration handpiece}} & \rot{\parbox[*][7ex][c]{6cm}{phacoemulsifier \\handpiece}} & \rot{\parbox[*][7ex][c]{6cm}{implant \\injector}} & \rot{\parbox[*][7ex][c]{6cm}{primary \\incision knife}} & \rot{\parbox[*][7ex][c]{6cm}{secondary \\incision knife}} & \rot{\parbox[*][7ex][c]{6cm}{micromanipulator}} & \rot{\parbox[*][7ex][c]{6cm}{suture needle}} & \rot{\parbox[*][7ex][c]{6cm}{Mendez ring} } & \rot{\parbox[*][7ex][c]{6cm}{Mendez ring \& \\biomarker}} & \rot{\parbox[*][7ex][c]{6cm}{Bonn forceps \& \\ secondary \\incision knife}} & \rot{\parbox[*][7ex][c]{6cm}{primary \\incision knife \& \\Bonn forceps}} & \rot{\parbox[*][7ex][c]{6cm}{capsulorhexis \\cystotome \&\\ Bonn forceps}} & \rot{\parbox[*][7ex][c]{6cm}{phacoemulsifier \\handpiece \&\\ Bonn forceps}} & \rot{\parbox[*][7ex][c]{6cm}{phacoemulsifier \\handpiece \&\\ micromanipulator}} & \rot{\parbox[*][7ex][c]{6cm}{irrigation/aspiration \\handpiece \& \\ micromanipulator}} & \rot{\parbox[*][7ex][c]{6cm}{hydrodissection \\cannula \& \\ micromanipulator}} 
& \rot{\parbox[*][7ex][c]{6cm}{Troutman \\forceps \& \\ suture needle}} & \rot{\parbox[*][7ex][c]{6cm}{Recall}} \\
\hline
\end{tabular}
}
\label{APP:tab:conf_mat_LSTM_test}
\end{table*}

\begin{table*}[t]
\Huge
\def\arraystretch{1.5}
\centering
\caption{Confusion matrix rounded to 4 digits of the real test data set for the split trained CNN Classifier.}
\resizebox{\textwidth}{!}{%
\begin{tabular}{|c||c|c|c|c|c|c|c|c|c|c|c|c|c|c|c|c|c|c|c|c|c|c|c|c|c|c|c||c|}
\hline
\diagbox[dir=NW, width = 7.5cm] {\rule{0cm}{6.5cm}{\textbf{\Huge{Label}}}}{\raisebox{-0.5ex}{\textbf{\Huge{Prediction}}}} & \rot{\parbox[*][7ex][c]{6cm}{no tool in \\contact} } & \rot{\parbox[*][7ex][c]{6cm}{biomarker}} & \rot{\parbox[*][7ex][c]{6cm}{hydrodissection \\cannula}} & \rot{\parbox[*][7ex][c]{6cm}{Rycroft cannula}} & \rot{\parbox[*][7ex][c]{6cm}{viscoelastic \\cannula}} & \rot{\parbox[*][7ex][c]{6cm}{cotton}} & \rot{\parbox[*][7ex][c]{6cm}{capsulorhexis \\cystotome}} & \rot{\parbox[*][7ex][c]{6cm}{Bonn forceps}} & \rot{\parbox[*][7ex][c]{6cm}{capsulorhexis \\forceps}} & \rot{\parbox[*][7ex][c]{6cm}{Troutman \\forceps}} & \rot{\parbox[*][7ex][c]{6cm}{irrigation/aspiration handpiece}} & \rot{\parbox[*][7ex][c]{6cm}{phacoemulsifier \\handpiece}} & \rot{\parbox[*][7ex][c]{6cm}{implant \\injector}} & \rot{\parbox[*][7ex][c]{6cm}{primary \\incision knife}} & \rot{\parbox[*][7ex][c]{6cm}{secondary \\incision knife}} & \rot{\parbox[*][7ex][c]{6cm}{micromanipulator}} & \rot{\parbox[*][7ex][c]{6cm}{suture needle}} & \rot{\parbox[*][7ex][c]{6cm}{Mendez ring} } & \rot{\parbox[*][7ex][c]{6cm}{Mendez ring \& \\biomarker}} & \rot{\parbox[*][7ex][c]{6cm}{Bonn forceps \& \\ secondary \\incision knife}} & \rot{\parbox[*][7ex][c]{6cm}{primary \\incision knife \& \\Bonn forceps}} & \rot{\parbox[*][7ex][c]{6cm}{capsulorhexis \\cystotome \&\\ Bonn forceps}} & \rot{\parbox[*][7ex][c]{6cm}{phacoemulsifier \\handpiece \&\\ Bonn forceps}} & \rot{\parbox[*][7ex][c]{6cm}{phacoemulsifier \\handpiece \&\\ micromanipulator}} & \rot{\parbox[*][7ex][c]{6cm}{irrigation/aspiration \\handpiece \& \\ micromanipulator}} & \rot{\parbox[*][7ex][c]{6cm}{hydrodissection \\cannula \& \\ micromanipulator}} 
& \rot{\parbox[*][7ex][c]{6cm}{Troutman \\forceps \& \\ suture needle}} & \rot{\parbox[*][7ex][c]{6cm}{Recall}} \\
\hline
\end{tabular}%
}
\label{tab:conf_mat_CNN_split_test}
\end{table*}
\begin{table*}[t]
\Huge
\def\arraystretch{1.5}
\centering
\caption{Confusion matrix rounded to 4 digits of the real test data set for the split trained LSTM classifier.}
\resizebox{\textwidth}{!}{%
\begin{tabular}{|c||c|c|c|c|c|c|c|c|c|c|c|c|c|c|c|c|c|c|c|c|c|c|c|c|c|c|c||c|}
\hline
\diagbox[dir=NW, width = 7.5cm] {\rule{0cm}{6.5cm}{\textbf{\Huge{Label}}}}{\raisebox{-0.5ex}{\textbf{\Huge{Prediction}}}} & \rot{\parbox[*][7ex][c]{6cm}{no tool in \\contact} } & \rot{\parbox[*][7ex][c]{6cm}{biomarker}} & \rot{\parbox[*][7ex][c]{6cm}{hydrodissection \\cannula}} & \rot{\parbox[*][7ex][c]{6cm}{Rycroft cannula}} & \rot{\parbox[*][7ex][c]{6cm}{viscoelastic \\cannula}} & \rot{\parbox[*][7ex][c]{6cm}{cotton}} & \rot{\parbox[*][7ex][c]{6cm}{capsulorhexis \\cystotome}} & \rot{\parbox[*][7ex][c]{6cm}{Bonn forceps}} & \rot{\parbox[*][7ex][c]{6cm}{capsulorhexis \\forceps}} & \rot{\parbox[*][7ex][c]{6cm}{Troutman \\forceps}} & \rot{\parbox[*][7ex][c]{6cm}{irrigation/aspiration handpiece}} & \rot{\parbox[*][7ex][c]{6cm}{phacoemulsifier \\handpiece}} & \rot{\parbox[*][7ex][c]{6cm}{implant \\injector}} & \rot{\parbox[*][7ex][c]{6cm}{primary \\incision knife}} & \rot{\parbox[*][7ex][c]{6cm}{secondary \\incision knife}} & \rot{\parbox[*][7ex][c]{6cm}{micromanipulator}} & \rot{\parbox[*][7ex][c]{6cm}{suture needle}} & \rot{\parbox[*][7ex][c]{6cm}{Mendez ring} } & \rot{\parbox[*][7ex][c]{6cm}{Mendez ring \& \\biomarker}} & \rot{\parbox[*][7ex][c]{6cm}{Bonn forceps \& \\ secondary \\incision knife}} & \rot{\parbox[*][7ex][c]{6cm}{primary \\incision knife \& \\Bonn forceps}} & \rot{\parbox[*][7ex][c]{6cm}{capsulorhexis \\cystotome \&\\ Bonn forceps}} & \rot{\parbox[*][7ex][c]{6cm}{phacoemulsifier \\handpiece \&\\ Bonn forceps}} & \rot{\parbox[*][7ex][c]{6cm}{phacoemulsifier \\handpiece \&\\ micromanipulator}} & \rot{\parbox[*][7ex][c]{6cm}{irrigation/aspiration \\handpiece \& \\ micromanipulator}} & \rot{\parbox[*][7ex][c]{6cm}{hydrodissection \\cannula \& \\ micromanipulator}} 
& \rot{\parbox[*][7ex][c]{6cm}{Troutman \\forceps \& \\ suture needle}} & \rot{\parbox[*][7ex][c]{6cm}{Recall}} \\
\hline
\end{tabular}%
}
\label{tab:conf_mat_LSTM_split_test}
\end{table*}

\clearpage

\twocolumn
\bibliographystyle{model2-names.bst}\biboptions{authoryear}
\bibliography{refs}

\end{document}